\documentclass[aps,prb,reprint,superscriptaddress]{revtex4-1}
\usepackage{graphicx}
\usepackage{fullpage}
\usepackage{amsmath,amssymb}
\usepackage{color}
\usepackage{float}
\usepackage{setspace}
\usepackage{hyperref}
\usepackage{array}
\usepackage[utf8]{inputenc}
\usepackage[export]{adjustbox}

 \makeatother

\begin{document}

\title{Epitaxial electrical contact to graphene on SiC}

%
%
\author{T. Le Quang}
\affiliation{Univ. Grenoble Alpes/CEA, INAC-PHELIQS, F-38000 Grenoble, France}

\author{ L. Huder}
\affiliation{Univ. Grenoble Alpes/CEA, INAC-PHELIQS, F-38000 Grenoble, France}

\author{ F. Lipp Bregolin}
\affiliation{Univ. Grenoble Alpes/CEA, INAC-PHELIQS, F-38000 Grenoble, France}

\author{A. Artaud}
\affiliation{Univ. Grenoble Alpes/CEA, INAC-PHELIQS, F-38000 Grenoble, France}
\affiliation{Univ. Grenoble Alpes/CNRS, Inst. NEEL, F-38000 Grenoble, France}

\author{H. Okuno}
\affiliation{Univ. Grenoble Alpes/CEA, INAC-MEM, F-38000 Grenoble, France}

\author{S. Pouget}
\affiliation{Univ. Grenoble Alpes/CEA, INAC-MEM, F-38000 Grenoble, France}

\author{N. Mollard}
\affiliation{Univ. Grenoble Alpes/CEA, INAC-MEM, F-38000 Grenoble, France}

\author{G. Lapertot}
\affiliation{Univ. Grenoble Alpes/CEA, INAC-PHELIQS, F-38000 Grenoble, France}

\author{A. G. M Jansen}
\affiliation{Univ. Grenoble Alpes/CEA, INAC-PHELIQS, F-38000 Grenoble, France}

\author{F. Lefloch}
\affiliation{Univ. Grenoble Alpes/CEA, INAC-PHELIQS, F-38000 Grenoble, France}
 
\author{E. F. C. Driessen}
\affiliation{IRAM, Institut de Radioastronomie Millimétrique St. Martin d’Hères France} 

\author{C. Chapelier}
\affiliation{Univ. Grenoble Alpes/CEA, INAC-PHELIQS, F-38000 Grenoble, France}

\author{V. T. Renard}
\affiliation{Univ. Grenoble Alpes/CEA, INAC-PHELIQS, F-38000 Grenoble, France}
\email{email: vincent.renard@cea.fr}

\begin{abstract}
Establishing good electrical contacts to nanoscale devices is a major issue for modern technology and contacting 2D materials is no exception to the rule. One-dimensional edge-contacts to graphene were recently shown to outperform surface contacts but the method remains difficult to scale up.  We report a resist-free and scalable method to fabricate few graphene layers with electrical contacts in a single growth step. This method derives from the discovery reported here of the growth of few graphene layers on a metallic carbide by thermal annealing of a carbide forming metallic film on SiC in high vacuum. We exploit the combined effect of edge-contact and partially-covalent surface epitaxy between graphene and the metallic carbide to fabricate devices in which low contact-resistance and Josephson effect are observed. Implementing this approach could significantly simplify the realization of large-scale graphene circuits.

\end{abstract}

\maketitle 


\section{Introduction}
Surface electrical contacts to two-dimensional materials suffer from the poor coupling between the 2D surface and the 3D metal.\cite{Leonard2011,Xia2011,Allain2015} The situation is further degraded by contamination in the lithographic processing and/or layer transfer.\cite{Robinson2011}  The best innovation is the recent realization of one-dimensional edge-contacts to graphene\cite{Wang2013} possibly combined with large doping.\cite{Park2016} However, the improvement of contact resistance is made at the expense of technological simplicity since the contact fabrication necessitates several steps. Edge bonding\cite{Borovikov2009,Taisuke2010,Kusunoki2015} and large electron transfer\cite{Berger2004,Berger2006} are known to occur during the growth of graphene on SiC and could be exploited for electrical contacts if SiC was replaced by a similar yet conducting material. Conducting carbides appear as good candidates since they have similar chemical properties and since they could allow new functionalities owing to additional material properties such as magnetism or superconductivity.
The growth of graphene on carbides other than SiC was first demonstrated by Foster, Long and Strumpf.\cite{Foster1958} In 1958, they showed that ``aluminum carbide dissociates in the vicinity of 2200-2500 $^\circ$C, at atmospheric pressures, to aluminum vapor and pure single crystals of graphite''  establishing that other carbides could potentially be used for graphene technology. Nevertheless, this subject has remained unexplored owing to the lack of commercial substrates.\cite{Presser2011} In this work, we demonstrate that few graphene layer can be grown on a metallic carbide by thermal annealing of a carbide forming metal film (niobium or tantalum) on SiC in high vacuum circumventing the problem of metallic carbide substrate availability. Based on this discovery we describe a resist-free and scalable method to fabricate few graphene layers (FGL) with electrical contacts in a single growth step. The combined effect of edge-contact\cite{Wang2013} and partially-covalent surface epitaxy\cite{Aizawa1992,Hwang1992} between graphene and the metallic carbide allows us to fabricate devices in which low contact-resistance and Josephson effect are observed.

\begin{figure*}
\centering
\large\raisebox{3.6cm}{a)}~\includegraphics[height=0.5\columnwidth]{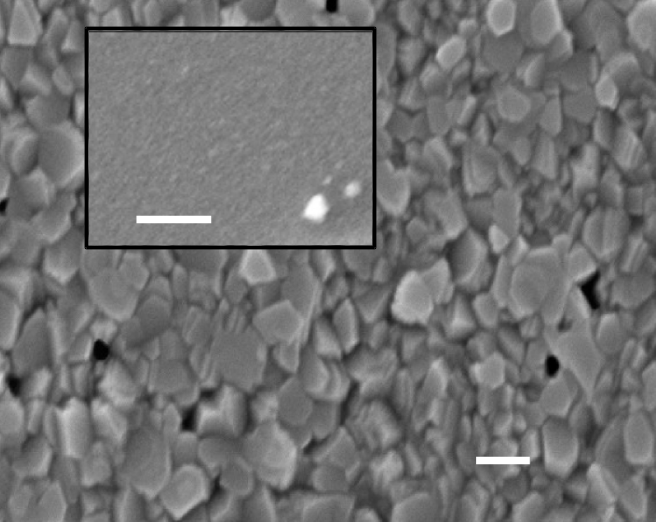}
\large\raisebox{3.6cm}{b)}~\raisebox{-.2cm}{\includegraphics[height=.5\columnwidth]{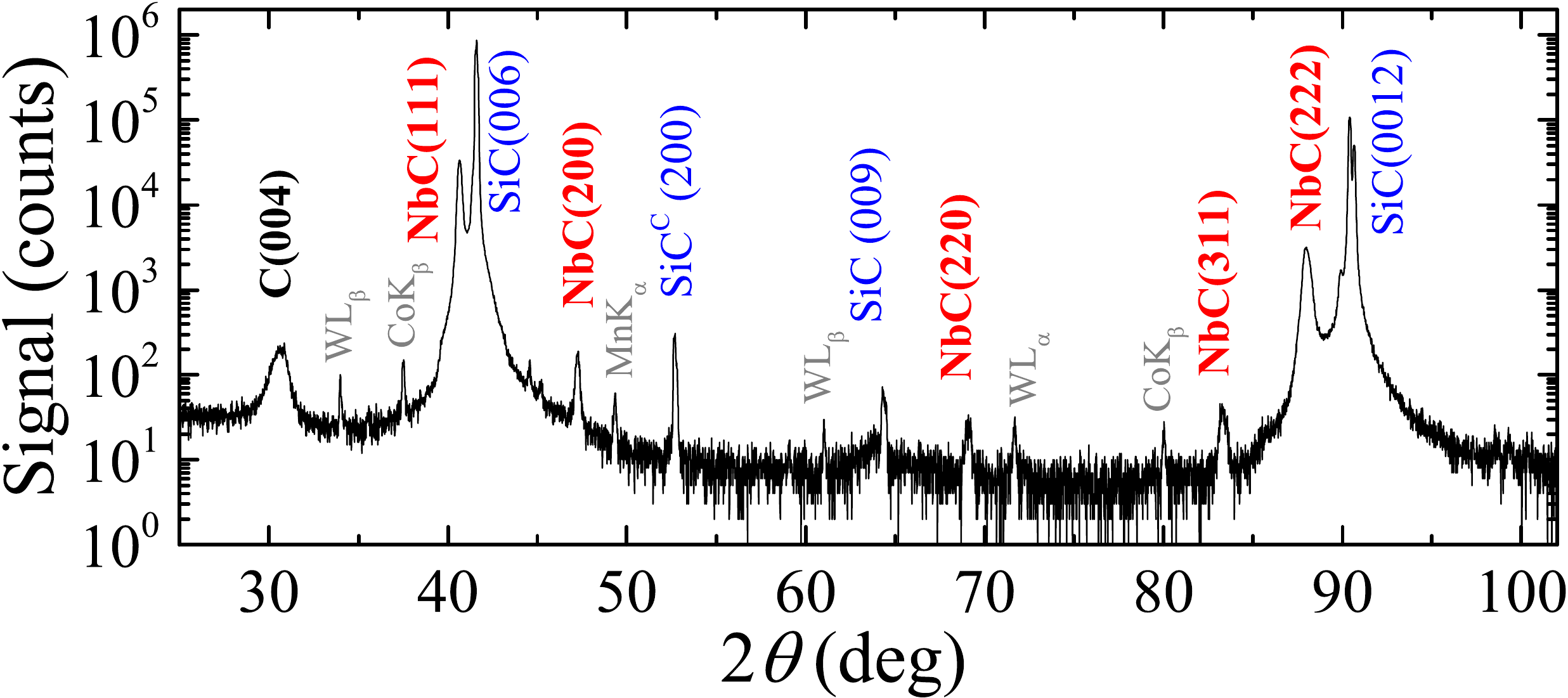}}
\large\raisebox{3.6cm}{c)}~\raisebox{-.2cm}{\includegraphics[height=.52\columnwidth]{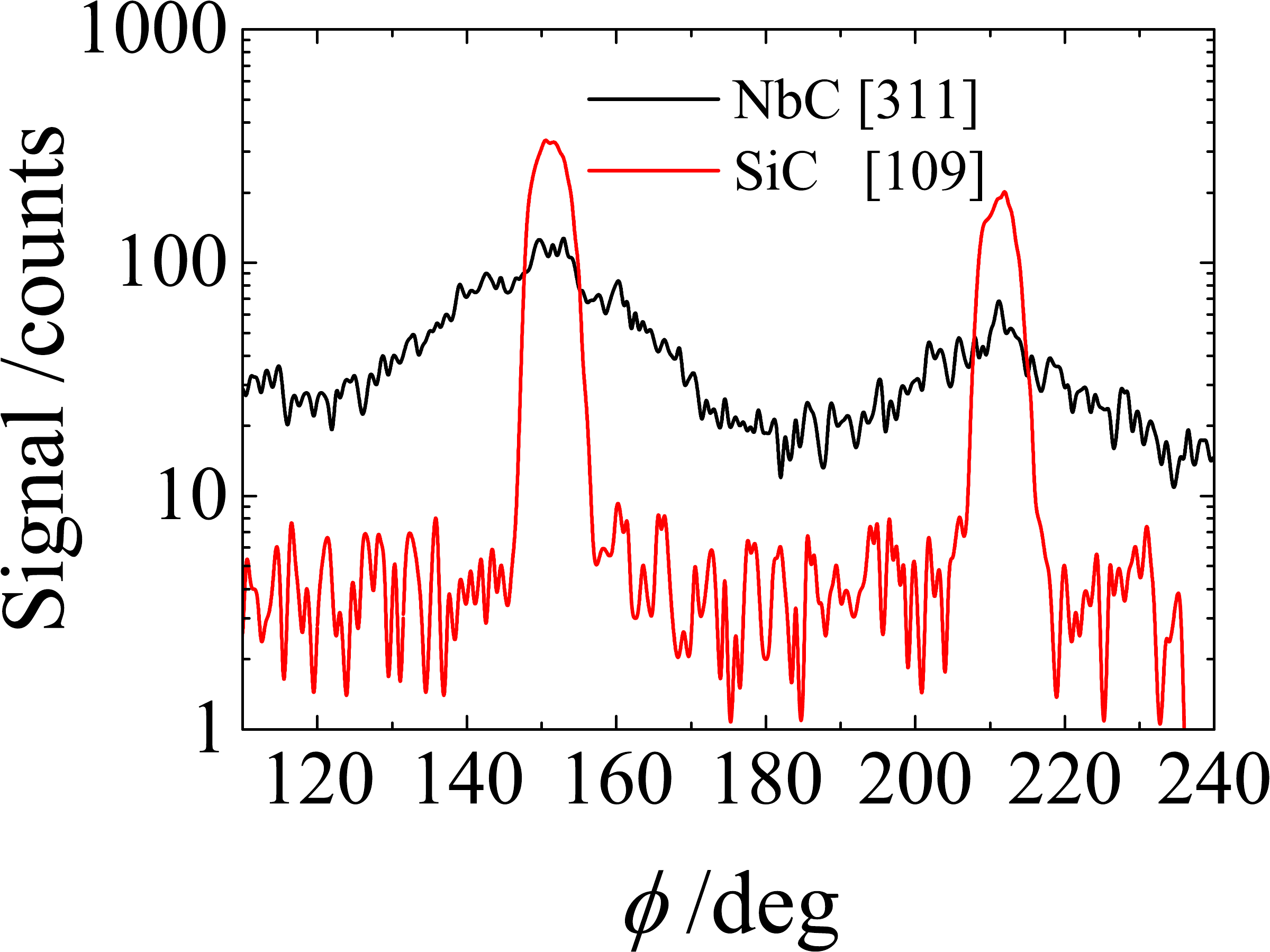}}
\large\raisebox{3.6cm}{d)}~\includegraphics[height=0.5\columnwidth]{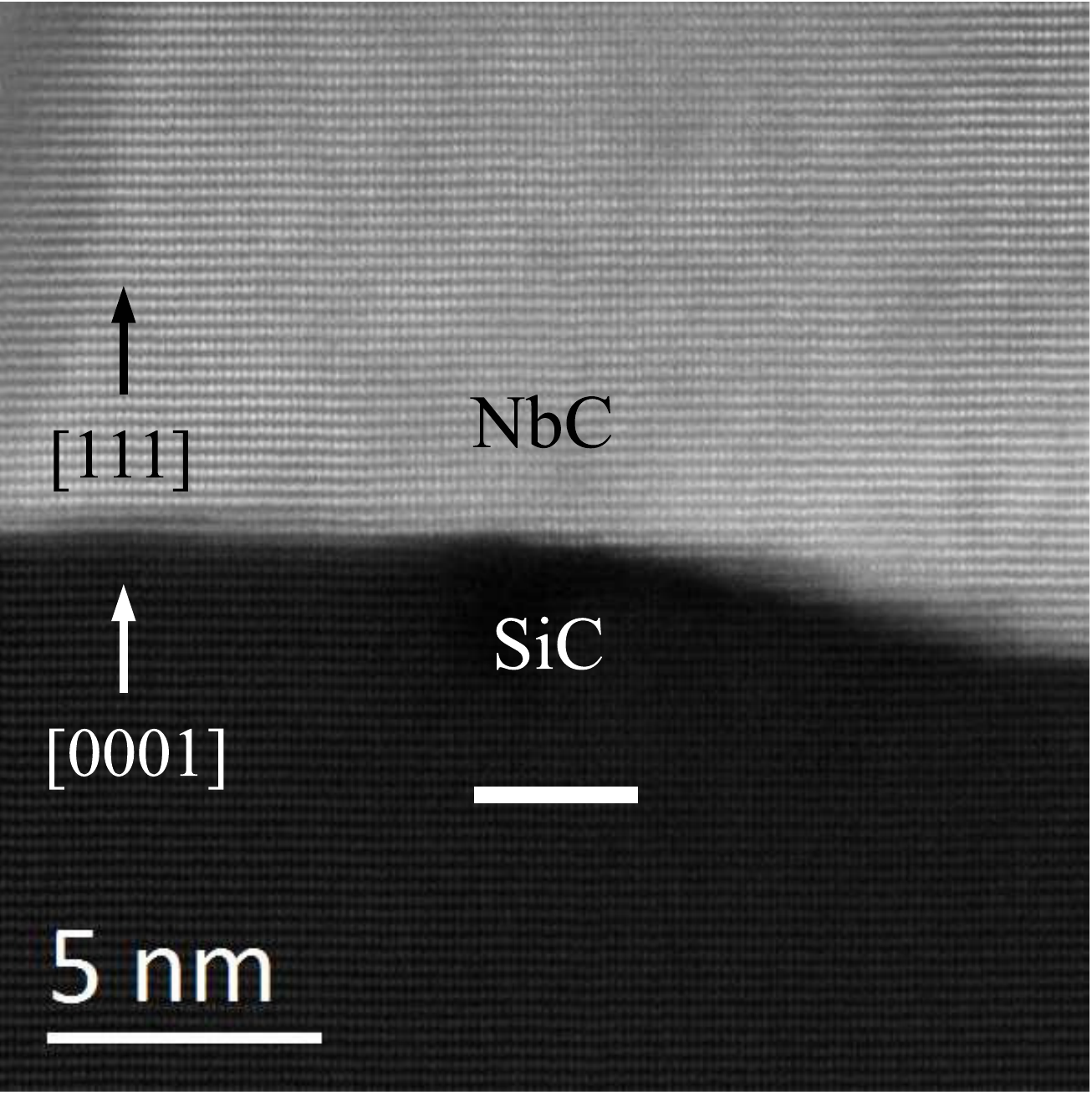}
\caption{a) Scanning electron microscope image of the niobium layer after thermal annealing  at 1360$^\circ$C. The inset shows a SEM image of the film prior to the annealing. Both scale bars are 200 nm. b) XRD pattern measured on the annealed film. The peaks of NbC are labeled in red, that of graphite in black, those of SiC in blue and those of experimental artifacts in grey. c) $\phi$-scan of the (311) and (109) reflections of respectively NbC and SiC. d) High resolution STEM image of the interface showing that NbC crystal grows in registry with the SiC substrate. The scale bar is 2.5 nm. \label{Fig1}}
\end{figure*}

\section{Experimental procedure}

\subsection{Sample preparation}

The NbC films were prepared by depositing a 40~nm thick layer of niobium by e-beam evaporation or magnetron sputtering on top of the carbon and silicon terminated surface of 4H-SiC substrates. The samples were then annealed in a RF-induction furnace inside a graphite crucible under a pressure lower than $10^{-6}$~mbar, following a recipe similar to that for graphene growth on SiC\cite{Berger2004}. The first annealing step ramps the temperature to 1140$^\circ$C in 60 minutes and holds this temperature for 30~minutes for degassing. The next temperature ramp is done in 180~minutes up to 1360$^\circ$C and temperature is then kept stable for 18 minutes. The cooling down to room temperature follows a reversed procedure.

\subsection{Characterization}

After the thermal treatment, the characteristics of the samples were investigated by several complementary techniques. The morphology was analyzed by scanning electron microscopy (SEM) and transmission electron microscopy (TEM). We used ZEISS Ultra+ SEM operated at 5 kV and a probe-aberration corrected FEI THEMIS (S)TEM operated at 200 kV.  A protective Ni layer was deposited before specimen preparation for TEM observations by focused ion beam.X-ray diffractograms (XRD) were measured using a PANalytical Empyrean diffractometer equipped with a Cobalt anode ($\lambda_{K_{\alpha 1}}$= $1.7890$ \AA, $\lambda_{K_{\alpha 2}}$ = 1.7929 \AA) with an Fe filter to reduce the $K_\beta$ contamination, a G\"{o}bel mirror and a 2D-Pixcel detector. For the Raman measurements, a circularly polarized Ar-laser ($\lambda$=$514.5~\mathrm{nm}$) at 1~mW of power was used as probe and the scattered light was dispersed by a Jobin-Yvon T64000 spectrometer and collected by a CCD detector. The spatial resolution was better than$~1~\mathrm{\mu m}$, and the spectral resolution was about 1 cm$^{-1}$.
Electrical measurements of the superconducting properties of the carbides were performed down to 1 K in a Quantum Design PPMS cryostat using a van der Pauw configuration with a $100~\mu$A drive current. The Josephson junctions measurements were done in a Cryoconcept dilution fridge using phase sensitive detection techniques.

\section{Results and discussion}
\subsection{Formation of the metallic carbide.}
We have formed a metallic carbide substrate by vacuum thermal annealing at 1360~$^\circ$C of a 40 nm thick niobium layer on the carbon face of SiC (See Methods). Figure~\ref{Fig1}a shows a scanning electron microscope (SEM) image of the Nb film before and after thermal annealing. While the pristine Nb film is flat, the annealed film becomes granular. Since Nb has a very high melting temperature (2477~$^\circ$C), we expect that this change in morphology is due to a solid state reaction between Nb and the SiC substrate  \cite{Burykina1968,Yaney1990,Chou1990,Wang2009} rather than to the melting of the Nb film. The high angle X-ray reflectivity spectra (XRD) shown in Fig.~\ref{Fig1}b confirms that the film has undergone profound changes. This $\theta/2\theta$ scan of the annealed sample, reveals that besides the expected peaks of the SiC substrate, one observes peaks at 40.63$^\circ$ and 87.9$^\circ$ which we attribute to reflections from crystalline NbC. The relative intensities of the different NbC peaks do not match the expected ones for a randomly oriented polycrystalline phase. They indicate the coexistence of a main [111] textured part with a minor polycrystalline contribution. The $\phi$ azimuthal scans of the NbC(311) and SiC(109) (Fig.~\ref{Fig1}c) reflections show that the textured part of the NbC layer grows with the preferential in-plane orientations NbC(111)[$\overline{1}$11] parallel to SiC(0001)[10$\overline{1}$0] and NbC(111)[$\overline{2}$11] parallel to SiC(0001)[01$\overline{1}$0]. XRD results are confirmed by TEM analysis which reveal a clean SiC/NbC interface with orientation correlation between NbC and SiC lattices (Fig.~\ref{Fig1}d). We have observed in most cases NbC [111] oriented parallel to SiC [0001] with several in-plane orientations.
XRD indicates a complete transformation of Nb to NbC since neither pure Nb, nor other niobium carbide or silicide were observed. Besides, XRD reveals the presence of cubic silicon carbide which has already been reported at the interface between NbC and SiC.\cite{Yaney1990}

\subsection{Electrical characterization of the carbide} 

\begin{figure} 
 \centering
 \includegraphics[width=0.9\columnwidth]{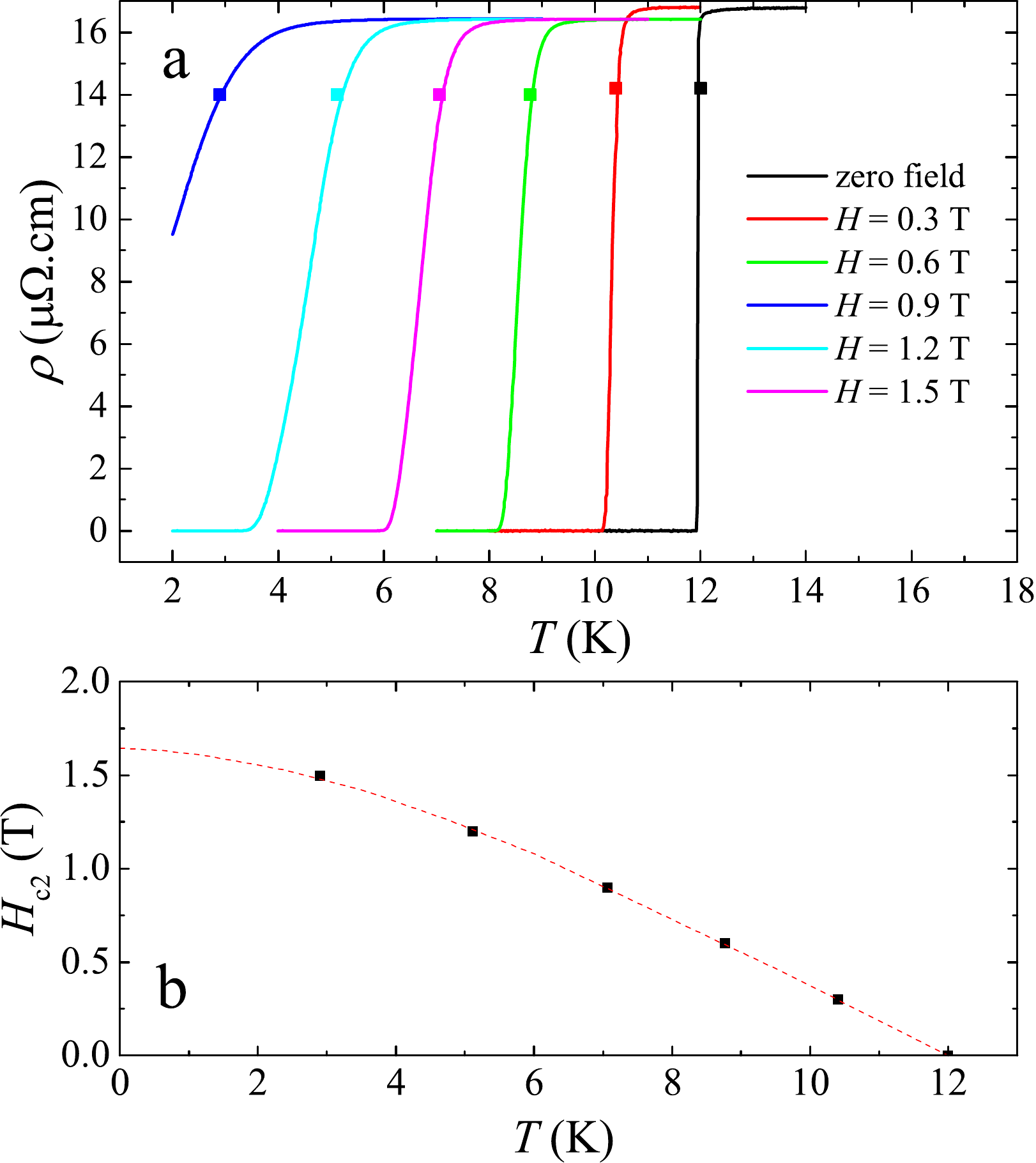}
 \caption{ a) Temperature dependence of the resistivity for different values of applied perpendicular magnetic field. Bullets represent the temperature for which $\rho$ is 80 \% of its normal state value. b) Temperature dependence of the upper critical magnetic field.  \label{Supra}}  
\end{figure}

We have measured the van der Pauw resistance of the NbC film studying all possible configurations for current injection and voltage probe contacts\cite{Vanderpauw1958}. This leads to an accurate estimation of the sheet resistance $R_s=3.0\pm 0.3~\Omega$. One then needs to know the film thickness $t$ to determine the film resistivity $\rho=R_s t$. From the TEM images we find that $t=55\pm10$ nm leading to $\rho=16\pm5~\mu\Omega.cm$. 
The diffusivity $D$ is obtained independently from the superconducting properties of the film. Figure~\ref{Supra}a shows the temperature dependence of the electrical resistivity of the sample for various applied magnetic fields. At zero field, the superconducting transition is sharp and occurs at 12~K. Under external fields, this transition becomes broadened and is shifted to lower temperatures. Figure~\ref{Supra}b shows the temperature dependence of the upper critical field defined as the field where the resistivity of the film is about 80$\%$ of its value in the normal state. Using the Werthamer, Helfand, Hohenberg dependence of the upper critical field $H_{c2}(T)$ for orbital pair breaking\cite{Werthamer1966} (red dashed line in Fig.~\ref{Supra}b), we estimate the critical field of our material to $H_{c2}$=1.64 T at T=0~K.
The diffusivity depends on the value of the slope of $H_{c2}(T)$ according to: $D=(4k_B /\pi e)(dH_{c2}/dT)^{-1}_{|T=T_c}$~\cite{Karasik1996}. We find that the electronic diffusivity of our NbC film is 7.3~$cm^{2}$/s. These transport properties can be related to the good stoichiometry  of the film which will be discussed in section III.D. They compete with those of the best NbC reported \cite{Golovashkin1986} and the method described here to synthesize NbC could prove useful for the realization of hot electron bolometers where large diffusivity and reasonably large RF impedance is needed.\cite{Karasik1996}

\begin{figure*}[t]
\centering
\large\raisebox{3.6cm}{a)}~\includegraphics[height=.5\columnwidth]{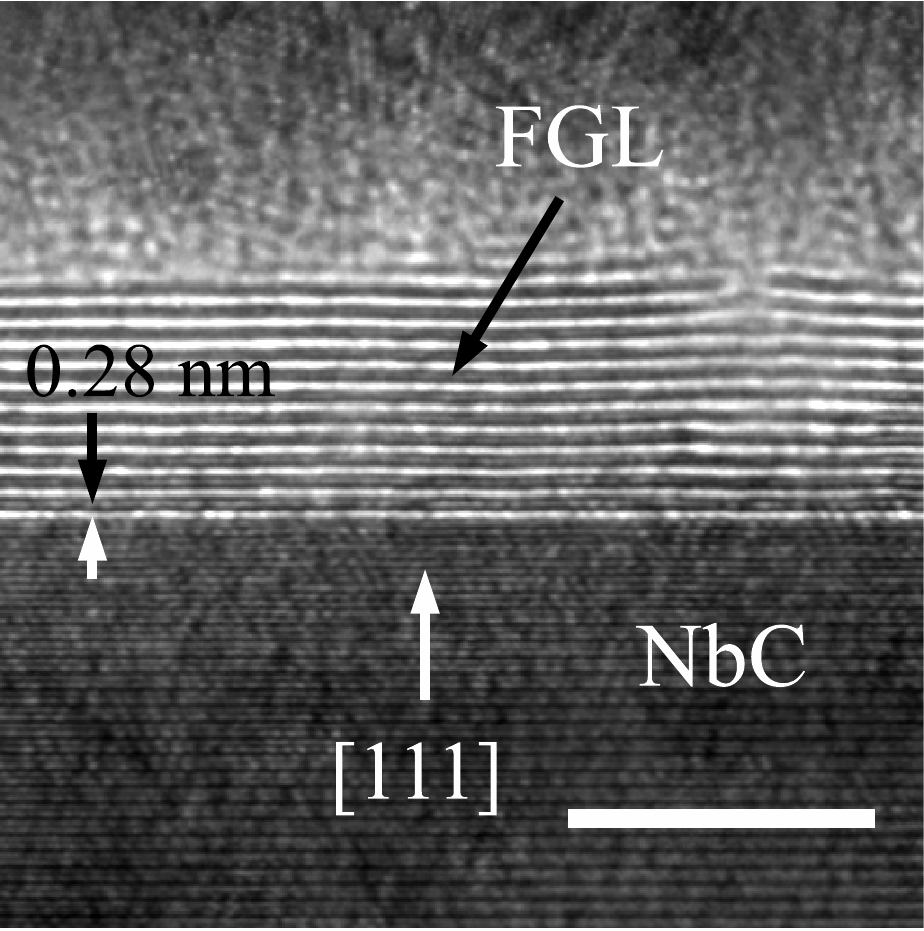}
\large\raisebox{3.6cm}{b)}~\includegraphics[height=.5\columnwidth]{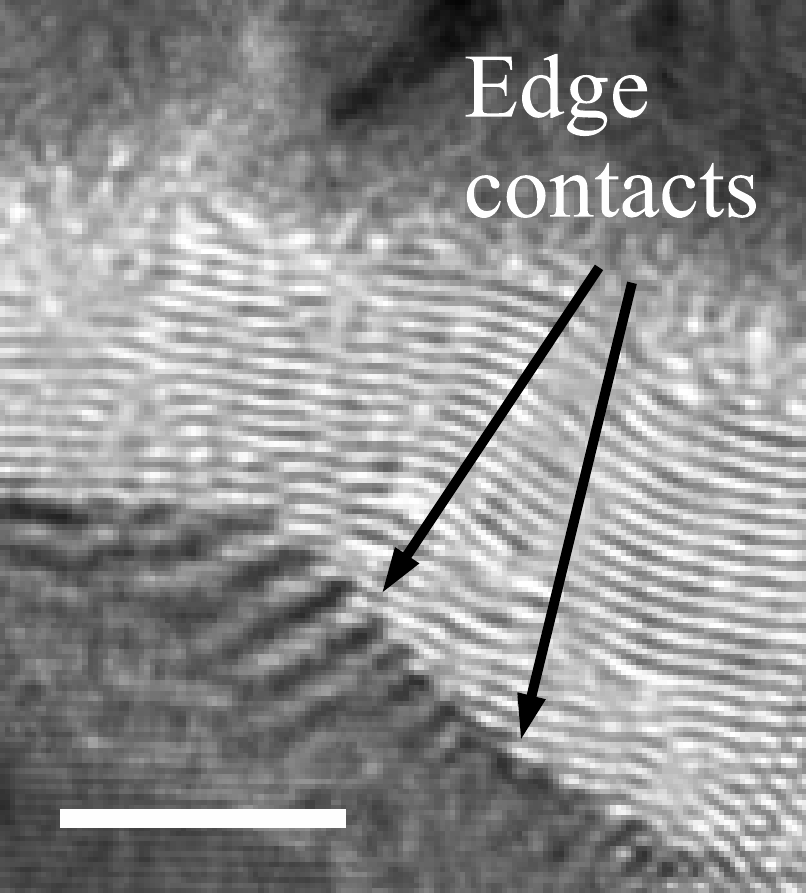}\\
\includegraphics[width=0.9\linewidth]{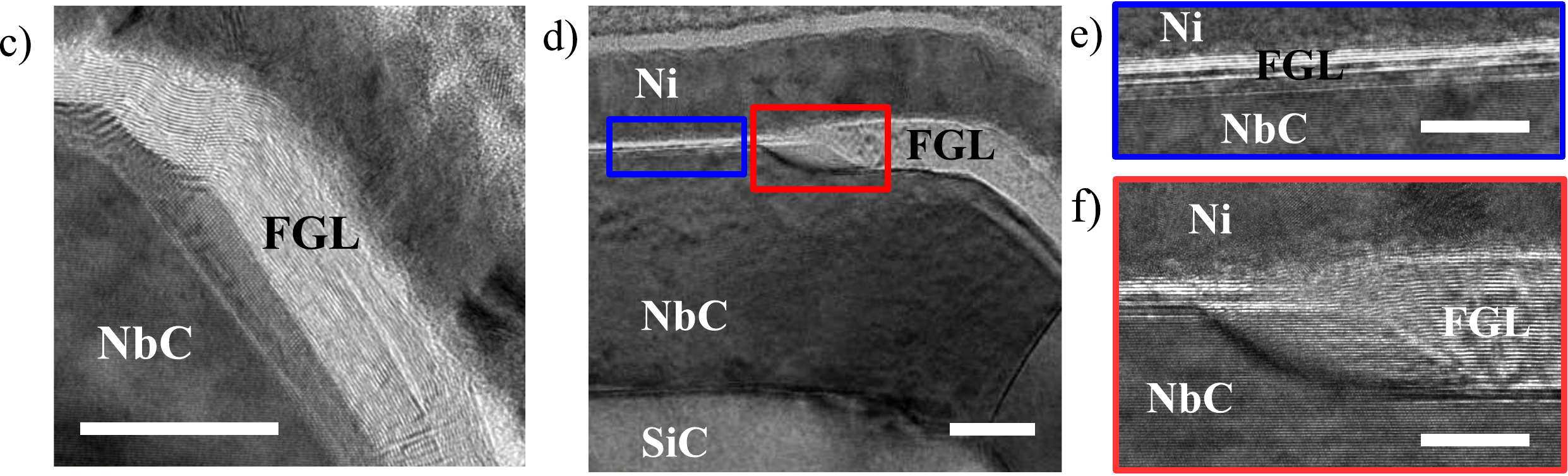}
\caption{ a) TEM image of NbC and the graphitic layer that is found on top of NbC. The scale bar is 5 nm. b) TEM image of the edge of a NbC grain showing graphene layers connected to the edge. The scale bar is 5 nm. c) TEM image showing a zoom out of panel a). The scale bar is 20 nm. d) TEM image of another NbC grain covered by FGL. The regions in the blue and red boxes are shown in e) and f). The scale bar is 20 nm in c) and 10 nm in e) and f). Note that a protective Ni layer was deposited before specimen preparation for TEM observations.
 \label{Fig:graphene}}
\end{figure*}
\subsection{Graphene on the metallic carbide.}

More interestingly, XRD reveals that the annealing process directly yields FGL at the surface of the NbC layer (XRD peak at 30.5$^\circ$). This is confirmed by the TEM image shown in Fig.~\ref{Fig:graphene}a where about ten layers of graphene are seen on top of NbC. The small distance between the first graphene layer and NbC (111) surface (0,28 nm, see Fig.~\ref{Fig:graphene}a) indicates a strong interaction between graphene and NbC.\cite{Aizawa1992} This interaction can be called partially covalent because its strength was shown to be between the weak van der Waals force and the normal covalent bond\cite{Aizawa1992} with strong electron transfer\cite{Hwang1992} for CVD graphene on metallic carbides. Figure~\ref{Fig:graphene}b shows that some graphene layers are bonded to a NbC terrace edge.

\subsection{Growth mechanisms.}

 Figure \ref{Fig:graphene}c and d show large field TEM views for two different NbC grains covered with FGL where details of the connection between NbC and FGL can be seen. The morphology observed in Fig.~\ref{Fig:graphene}d is very similar to that observed on graphene on SiC.\cite{Kimura2013} On the left of the image (blue box enlarged in Fig.~\ref{Fig:graphene}e) there are only few graphene layers while on the right of the image (red box enlarged in Fig.~\ref{Fig:graphene}f) there are much more graphene layers. On SiC, similar morphology was interpreted as resulting from the preferential dissociation of SiC at step edges which leads to nucleation and growth at step edges. This morphology suggests that the growth of graphene on NbC proceeds in a similar way as on SiC\cite{Borovikov2009,Taisuke2010,Kusunoki2015} . However, thermal decomposition of NbC at 1360~$^\circ$C raises questions since previous studies have revealed that NbC bulk single crystals do not dissociate to graphene by thermal annealing in vacuum at least up to 1800~$^\circ$C.\cite{Aizawa1992} Another scenario could be that SiC dissociates through the NbC film via diffusion in the bulk or at grain boundaries.
 
 \begin{figure*}[t]
\large\raisebox{3.8cm}{a)}\includegraphics[height=.26\linewidth]{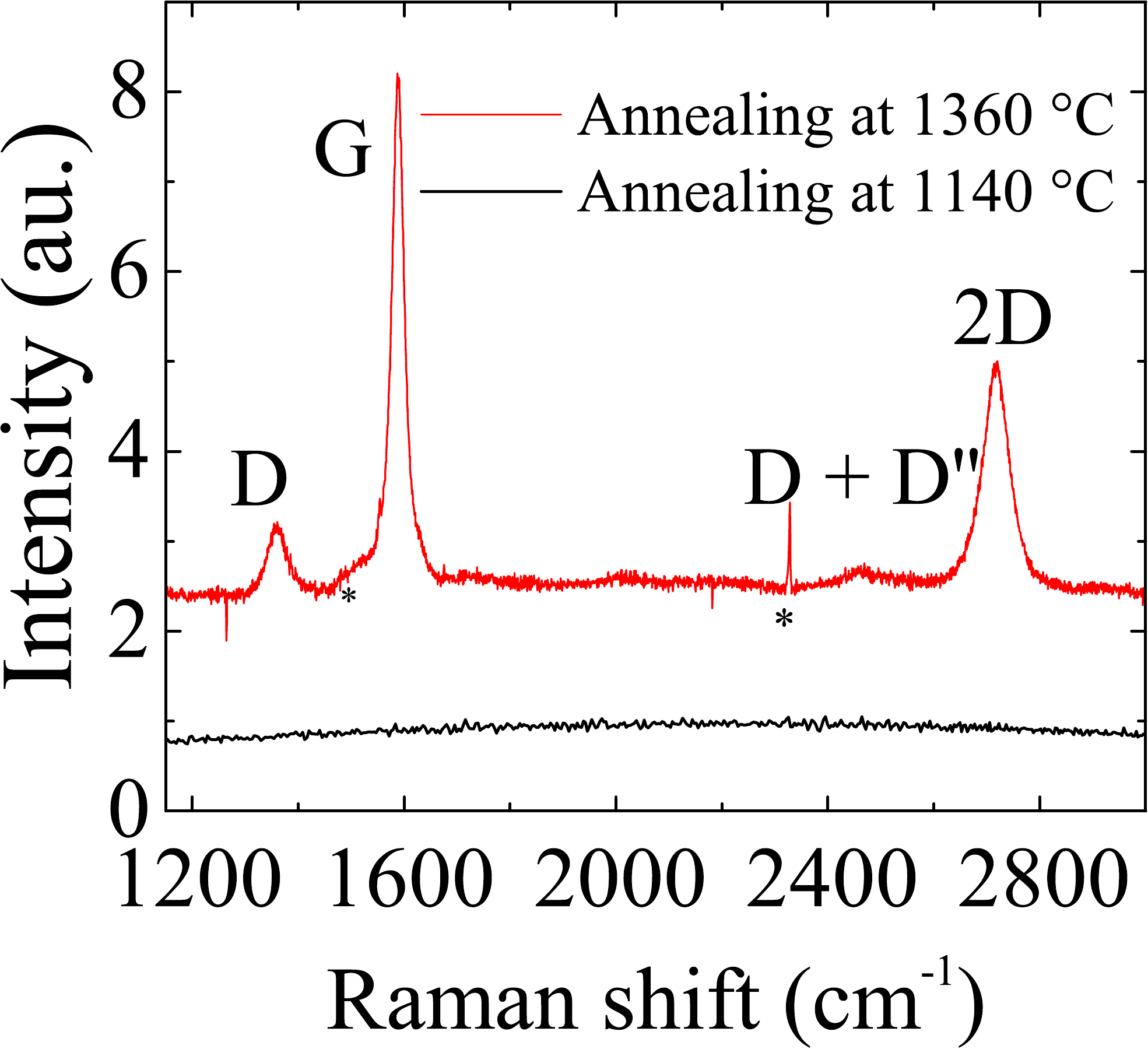}
\large\raisebox{3.8cm}{b)}\includegraphics[height=.26\linewidth]{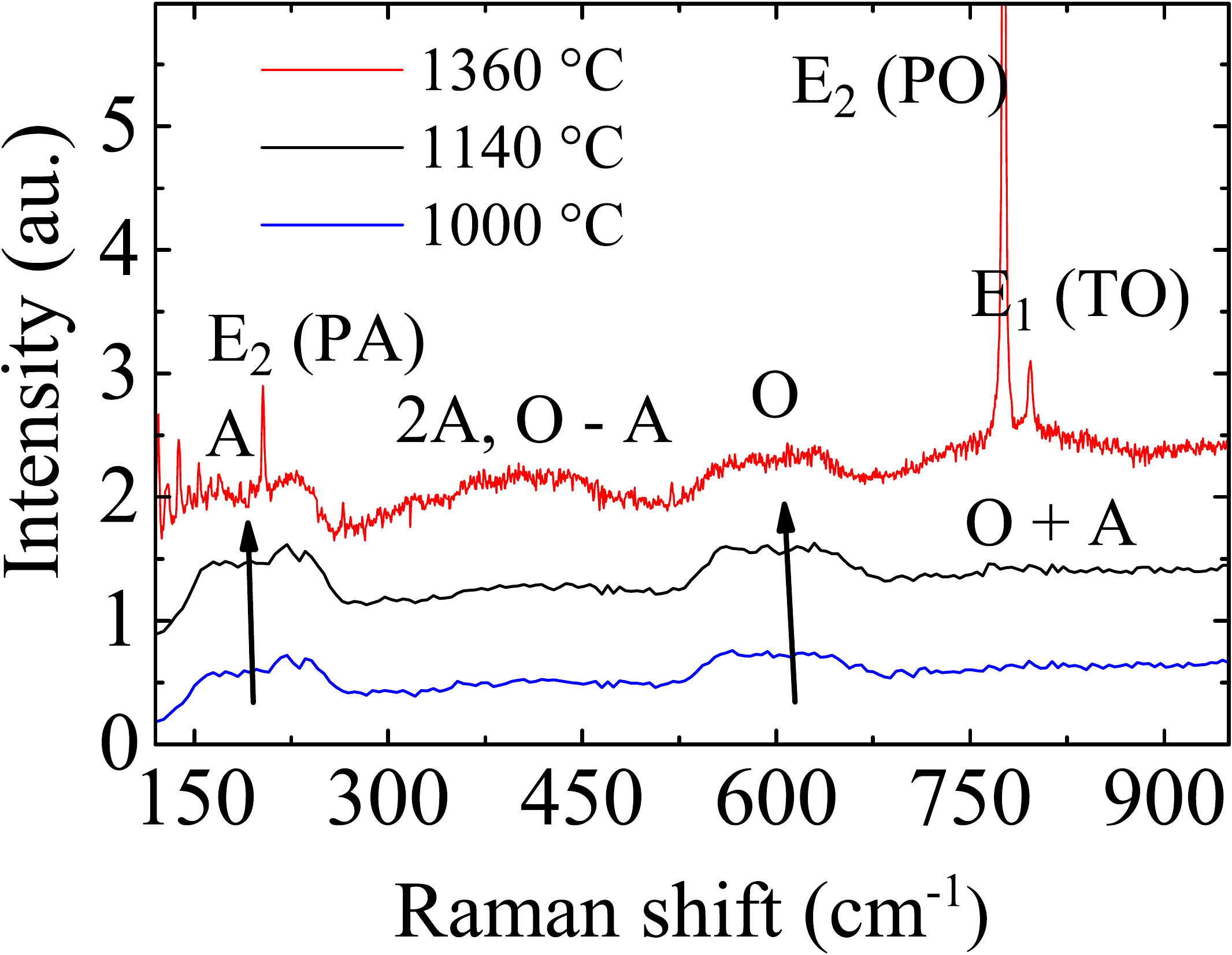}
\large\raisebox{3.8cm}{c)}\includegraphics[height=.26\linewidth]{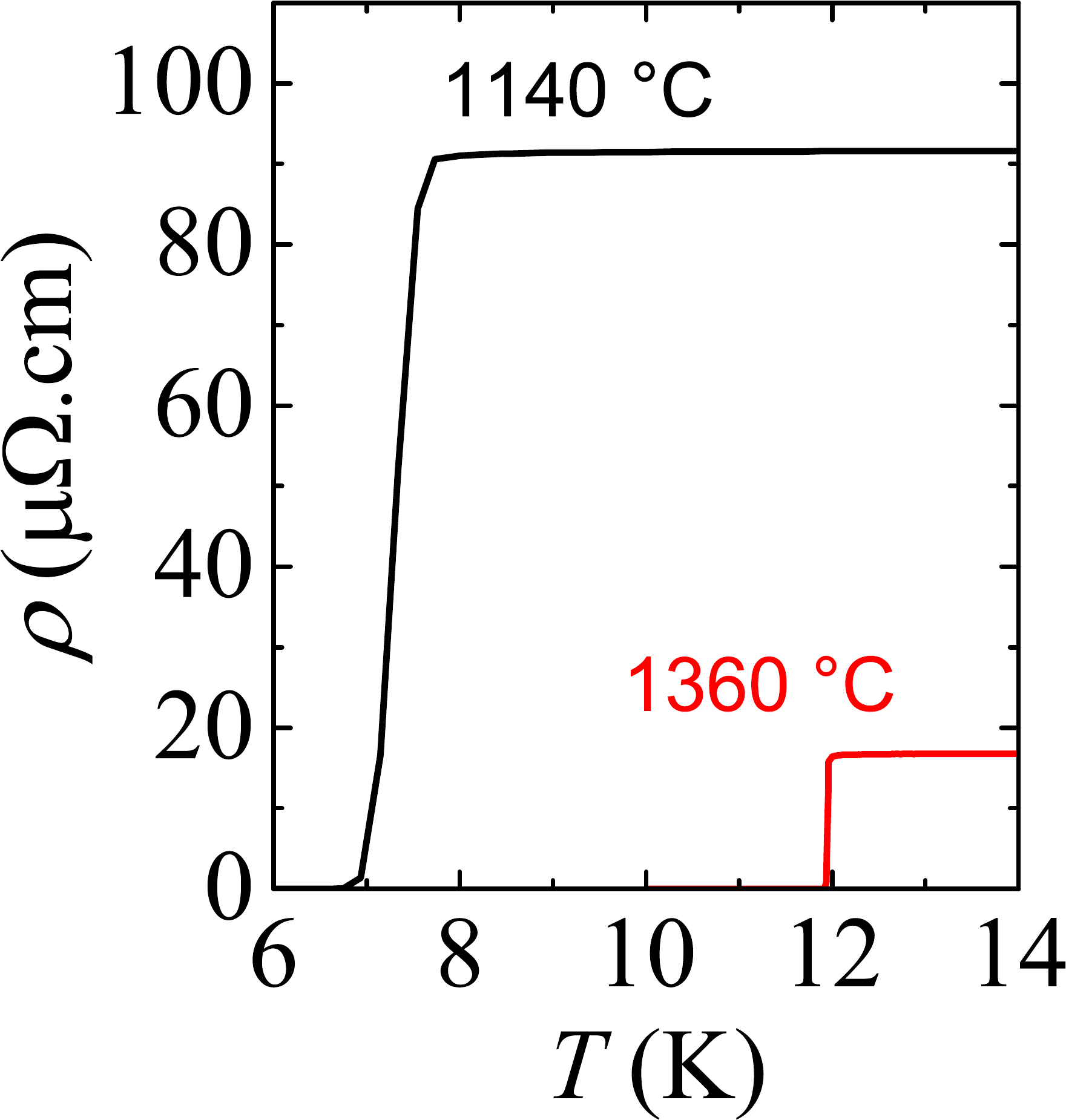}
\caption{a) The black line corresponds to the Raman spectrum measured on a sample annealed at 1140~$^\circ$C and is compared to that measured on a sample annealed at 1360~ $^\circ$C (red line). While the red spectrum displays typical sp$^2$ carbon signatures (G, D and 2D), those signatures are absent from the black spectrum indicating that no FGL are formed during the annealing at this temperature. Sharp peaks marked with asterisks are associated to the vibrational modes of O$_2$ and N$_2$ molecules. For clarity, the black spectrum has been shifted vertically. b) Low wavenumber Raman spectra acquired on samples annealed at different temperatures (red line 1360 $^\circ$C, black line 1140 $^\circ$C~ and blue line 1000 $^\circ$C). c) Low temperature resistivity of NbC films obtained by annealing at 1360$^\circ$C (red curve) and at 1140$^\circ$C (black curve).
 \label{Fig:Sequential_growth}}
\end{figure*}
 
Further experiments demonstrate that the carbidization of Nb and graphene growth occur sequentially rather than simultaneously. Indeed, no graphene signal is observed on the Raman spectrum of a Nb film annealed at 1140~$^\circ$C (Fig. \ref{Fig:Sequential_growth}a). Also, Raman measurements indicate that NbC grown at temperatures below 1360$^\circ$C is sub-stoichiometric. Figure \ref{Fig:Sequential_growth}b shows the low wave number Raman spectrum of three 40 nm Nb films after thermal annealing at 1360 $^\circ$C, 1140  $^\circ$C  and 1000 $^\circ$C~ respectively. These spectra show typical features of NbC\cite{wipf_vacancy-induced_1981}. Since NbC crystals inevitably contain C vacancies, quasi-momentum conservation is broken and both acoustic (A) and optical (O) phonons over the whole Brillouin zone are therefore Raman active. This results in doublets centered respectively on 170 and 230 cm$^{-1}$ (A), and 570 and 620 cm$^{-1}$ (O). We note that we did not find any signature of silicide or other carbide in the Raman response of our samples.
Acoustic (A) and optical (O) phonons are seen to shift to lower Raman shift as the annealing temperature increases. This global softening of NbC phonons can be interpreted as a lowering C vacancy concentration \cite{wipf_vacancy-induced_1981}. Moreover, additional combinations and overtones around 370 cm$^{-1}$ (2A,O-A) and 790 cm$^{-1}$ (O+A) develop when the Nb/C ratio approaches unity \cite{wipf_vacancy-induced_1981}. These resonances are observed in the sample annealed at 1360 $^\circ$C~ confirming our hypothesis that higher temperature annealing allows to saturate the carbide. Comparison with previous results \cite{wipf_vacancy-induced_1981} allows to set a lower limit of 0.98 to the carbon/niobium ratio in the sample annealed at 1360 $^\circ$C. This is further confirmed by the superconducting properties of the NbC films (Fig.~\ref{Fig:Sequential_growth}c). In NbC, $T_c$ is extremely sensitive to the carbon content and is degraded from above 12 K for perfect stoichiometry \cite{Dubistky2005} to less than 1.5 K for NbC$_{0.8}$.\cite{Giorgi1962} Figure~\ref{Fig:Sequential_growth}c shows that the resistivity vanishes for a critical temperature of 11.9 K for the sample annealed at 1360$^\circ$C while the critical temperature is only 7 K for the sample annealed at 1140$^\circ$C. Following Ref.~\cite{Golovashkin1986} in using the critical temperatures of Ref.~\cite{Giorgi1962} as a calibration we get NbC$_{0.98}$ and NbC$_{0.93}$ respectively, demonstrating that higher annealing temperature allows to saturate the carbide and eventually unlocks the growth of FGL. 

\begin{figure*}
\centering
\large\raisebox{3.2cm}{a)}~\includegraphics[height=0.5\columnwidth]{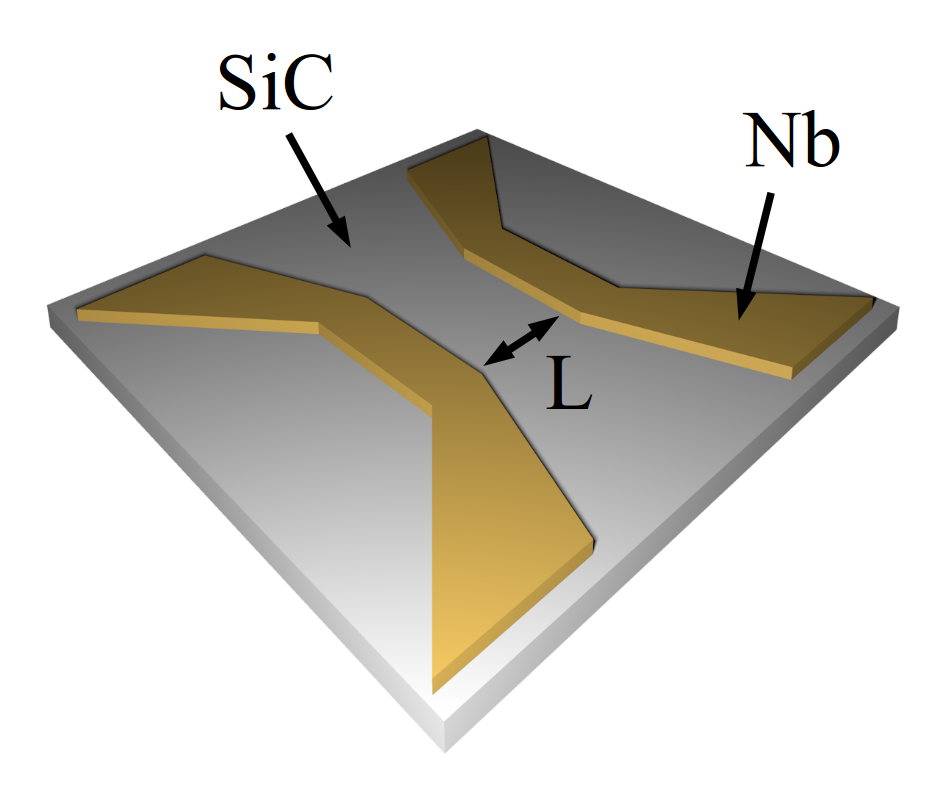}
\includegraphics[height=.5\columnwidth]{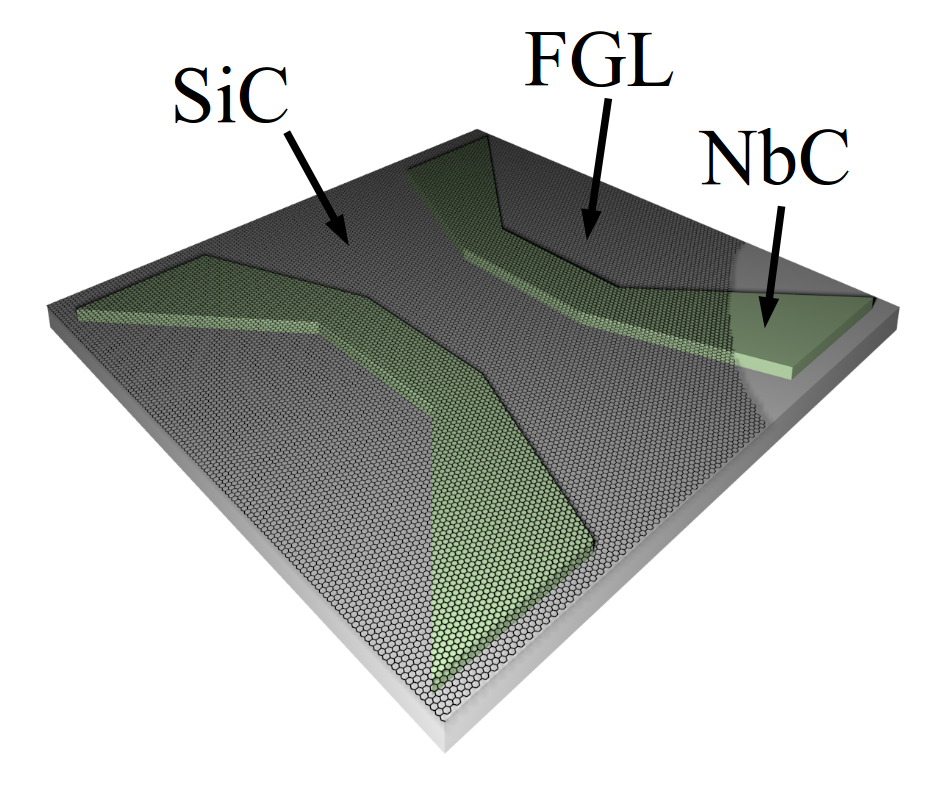}
\includegraphics[height=.5\columnwidth]{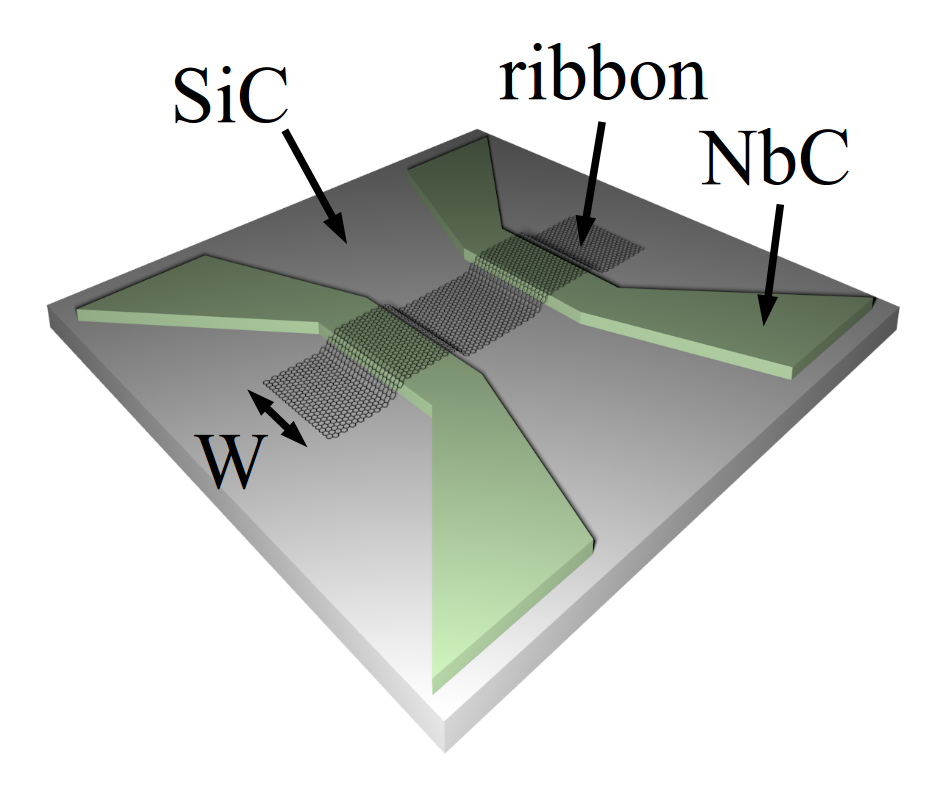}
\large\raisebox{5cm}{b)}\LARGE\raisebox{2.3cm}{c}~\raisebox{0.3cm}{\includegraphics[height=.55\columnwidth]{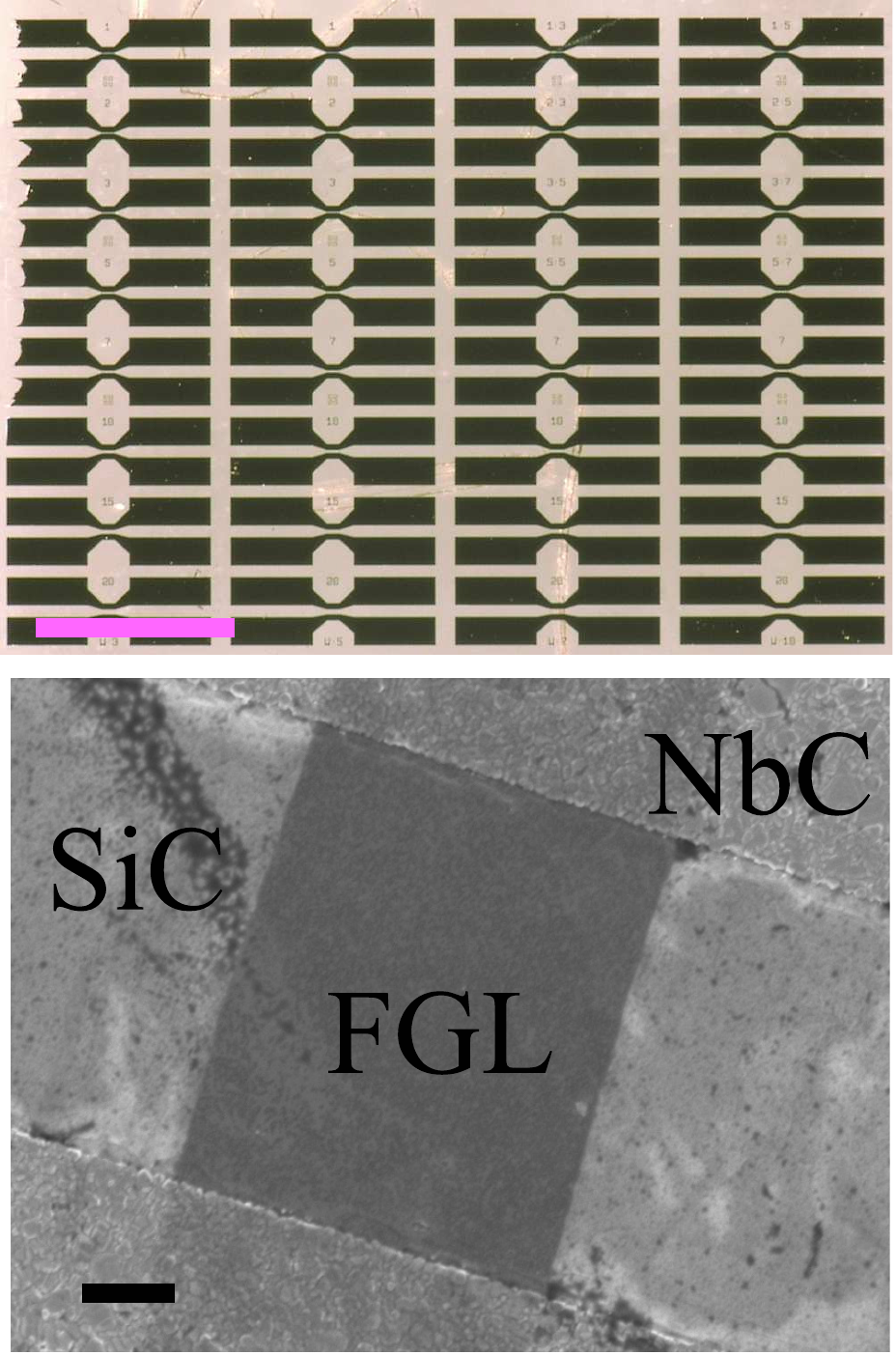}}
\large\raisebox{5cm}{d)}~\raisebox{0.3cm}{\includegraphics[height=0.55\columnwidth]{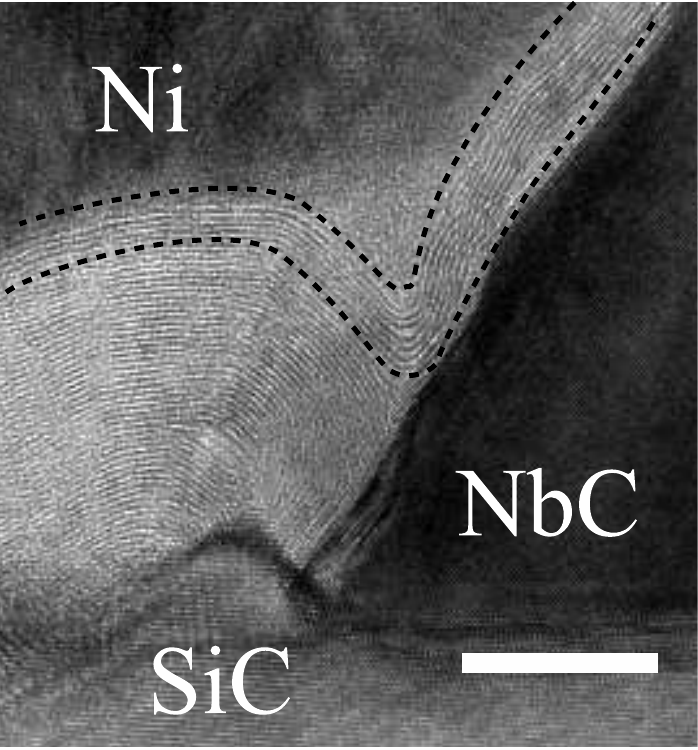}}
\large\raisebox{5cm}{e)}~\includegraphics[height=.55\columnwidth]{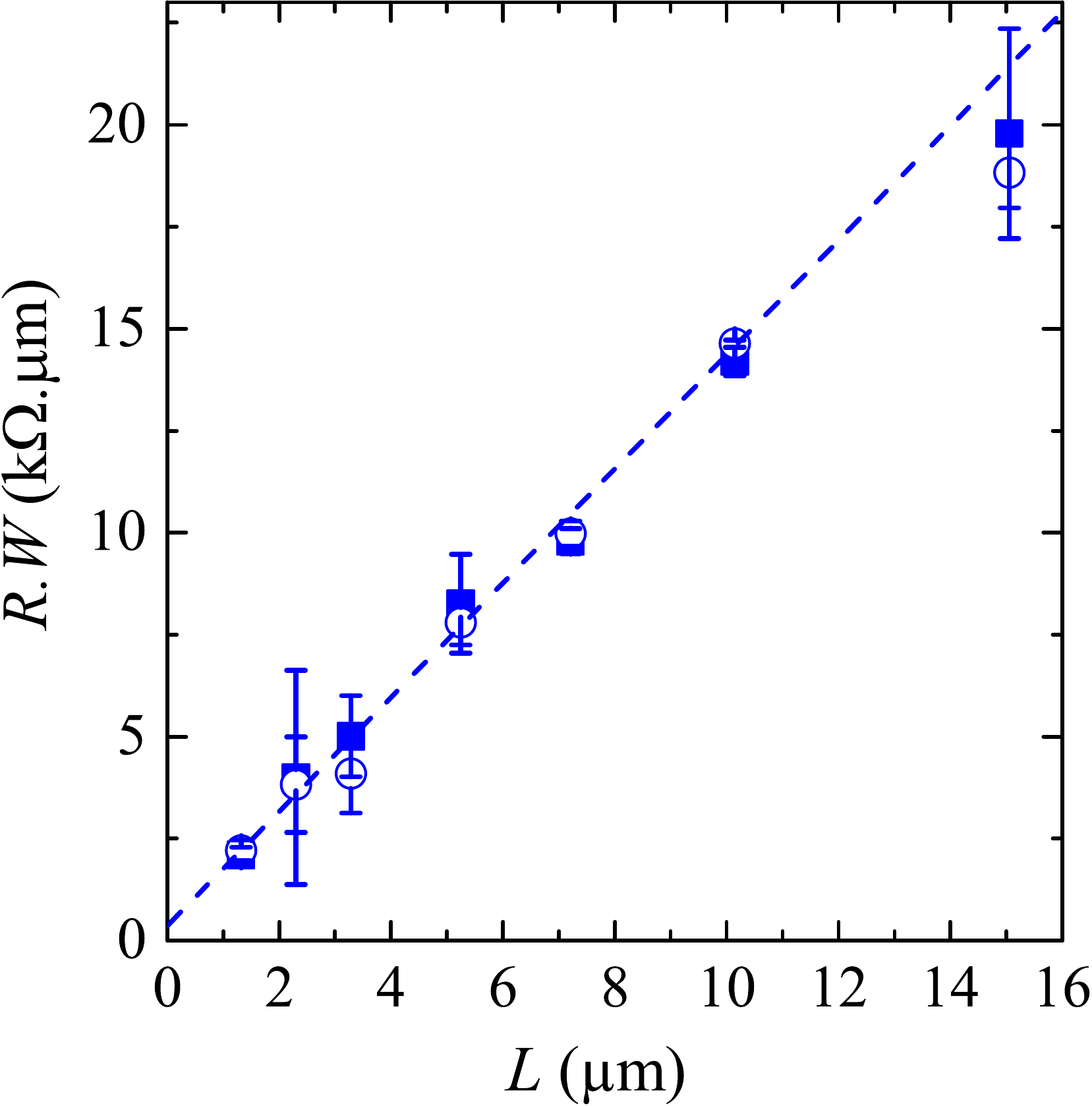}

\caption{a) The fabrication starts with the deposition of niobium electrodes on bare SiC (left). After high temperature annealing, the metallic electrodes have been transformed to NbC and the entire surface is covered with FGL (center). A ribbon is then etched in the FGL film using O$_2$ plasma (right). b) Optical microscope image of the chip with 32 four-terminal devices. The scale bar is 1 mm. c)  Scanning electron microscope image of a 5.2$\times$4.4~$\mu$m$^2$ ribbon. The scale bar is 1 $\mu$m. d) Transmission electron microscope image of the junction between NbC and SiC showing the FGL flowing continuously over the two materials. The dashed lines materialize continuous graphene layers at the junction between SiC and NbC. The scale bar is 10 nm. e) Resistance times width of the devices as function of the length. For each length 4 to 5 devices were used to determine the plotted mean value with standard deviation as an error bar. The resistance was measured at room temperature (\textcolor{blue}{$\bigcirc$}) and at 4~K (\textcolor{blue}{$\blacksquare$}). The expression $R.W=1400\times L+364$ $\Omega.\mu$m fits the experimental data (\textcolor{blue}{- - -}).\label{Fig2}}
\end{figure*}

\subsection{Fabrication of electrical devices and determination of the contact resistance}
Having demonstrated the growth of graphene on a metallic carbide, we have adapted the process to fabricate NbC/FGL/NbC devices where the electrical contact is established during the growth. The overall idea is to structure the initial layer of Nb and exploit the fact that the FGL grow continuously across the junction between SiC and NbC. The main steps of the device fabrication are described in Fig.~\ref{Fig2}a. Niobium leads are first deposited through a PMMA mask on the substrate at a distance $L=1-15~\mu$m defining the nominal junction length. The sample is then annealed at 1360~$^\circ$C in vacuum to form NbC and FGL which covers the entire sample. At the final step of the fabrication, ribbons of variable width $W=3-10~\mu$m are defined between the electrodes using O$_2$ plasma. Figure~\ref{Fig2}b shows an optical microscope image of a SiC chip with 32 devices demonstrating that the technique can be easily scaled up. Figure~\ref{Fig2}c shows a SEM image of a 5.2$\times4.4$~$\mu$m$^2$ device after the processing and the TEM image of Fig.~\ref{Fig2}d illustrates that the FGL film is continuous at the interface between NbC and SiC. The FGL film is thicker in the uncovered parts of SiC which further confirms that the growth of graphene on NbC is delayed by the carbidization of niobium.

\begin{figure}[t]
\large\raisebox{4cm}{a)}~\includegraphics[height=0.6\columnwidth, left]{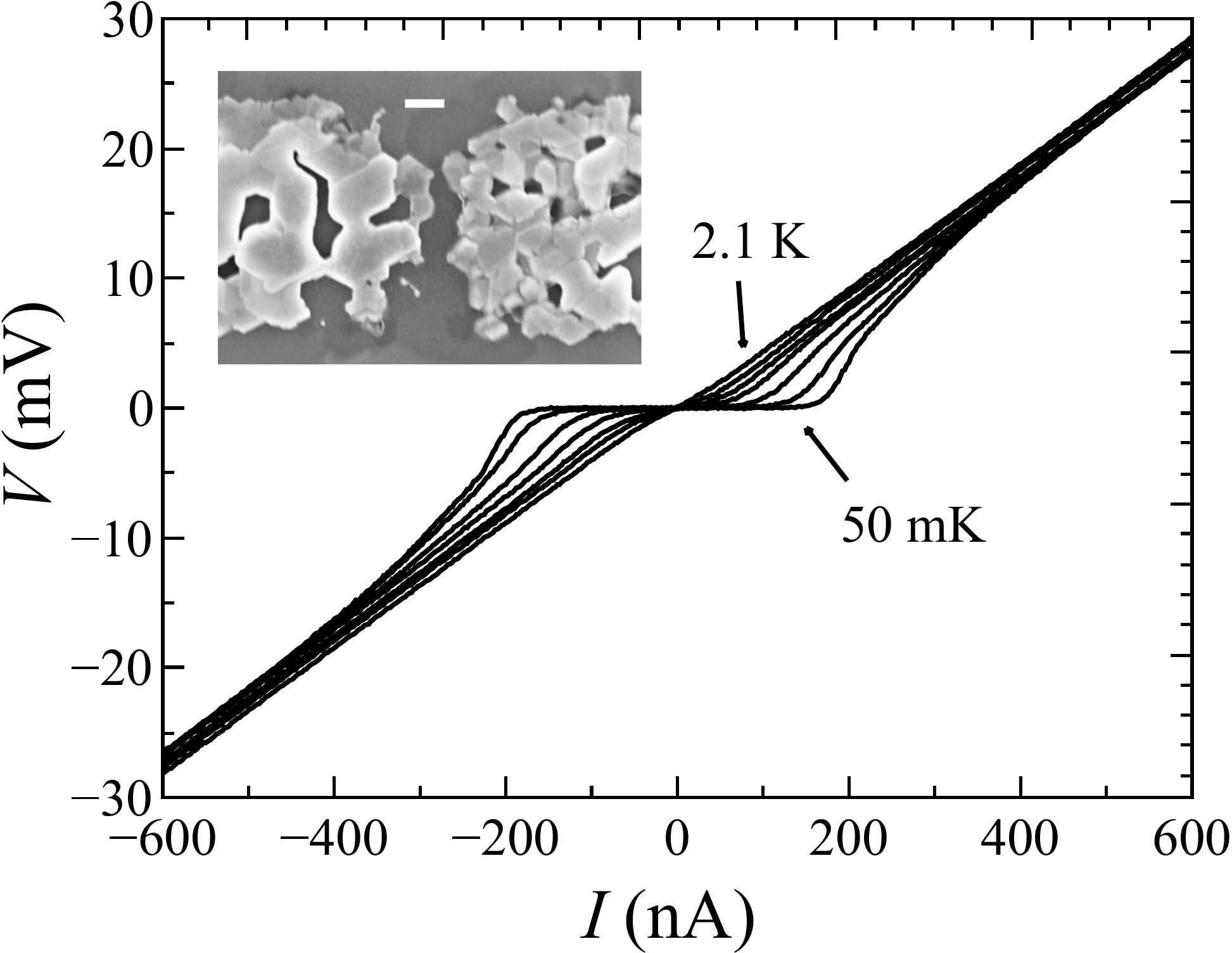}
\large\raisebox{4cm}{b)}~\includegraphics[height=0.6\columnwidth, left]{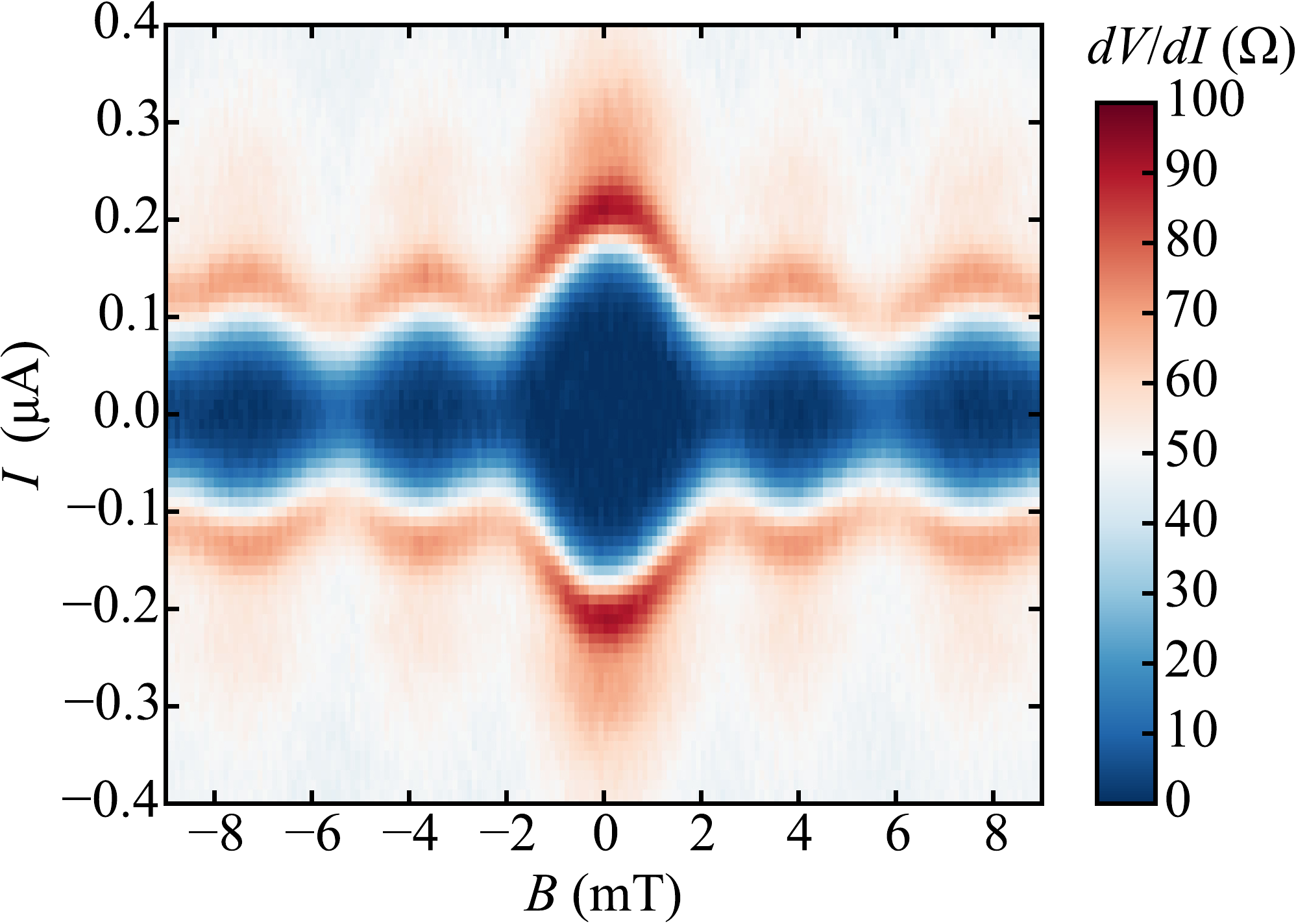}

\caption{\textbf{Characterization of a Josephson Junction.}  a) Temperature dependence of the current-voltage characteristic of the junction (T=50 mK, 300 mK, 600 mK, 900 mK, 1.2 K, 1.5 K, 2.1 K ). The inset presents a SEM image of the junction (the scale bar is 200 nm). b) Color plot of the magnetic field and current dependence of the junction's differential resistance at 50 mK.   \label{Fig3}}
\end{figure}
The devices of Fig.~\ref{Fig2}b have variable length and width to investigate the NbC/FGL contact resistance by the transfer length method.\cite{Berger1972} Figure~\ref{Fig2}e shows the product of the resistance of the devices and their width $R\times W$ as function of their length at room and liquid helium temperatures. Contrary to previously discussed specimens, the devices presented here were grown on the Si-face of the substrate which shows that the growth works similarly on both faces of SiC. As expected, the resistance is independent of temperature and proportional to the length. The extrapolation of the linear dependence to $L=0$~$\mu$m allows to estimate the specific contact resistance $2R_c\equiv R(L=0)\times W$. We find $R_c =182~\Omega.\mu m$ which equals the best reported contact resistance to multilayer graphene.\cite{Kazuyuki2015} We stress here that electrical contacts are realized during the growth which considerably simplifies the sample processing compared to other methods where the contacts are prepared later on.\cite{Chu2014,Kazuyuki2015} The low contact resistance results from three factors : First, the contact between the metal and graphene is exempt from contamination due to lithography. Indeed, despite resist is used in the initial deposition of Nb, any residues are blown away during the degassing at 1140 $^\circ$C (See Methods) leaving a clean NbC/graphene interface formed during the growth. Secondly, the peculiar texturation of NbC on SiC leads to a partially covalent epitaxy of graphene on NbC[111].\cite{Aizawa1992} Large electron transfer specific to this interface\cite{Hwang1992} was reported which, together with the low graphene/NbC distance lowers the contact resistance.\cite{Xia2011} Thirdly, Fig.~\ref{Fig:graphene}b and d shows that edge contacts\cite{Wang2013} are also formed during epitaxy which also improve the contact resistance to multilayer graphene.\cite{Chu2014,Kazuyuki2015}  

\begin{figure*}[t]

\large\raisebox{4cm}{a)}~\raisebox{.3cm}{\includegraphics[height=0.55\columnwidth]{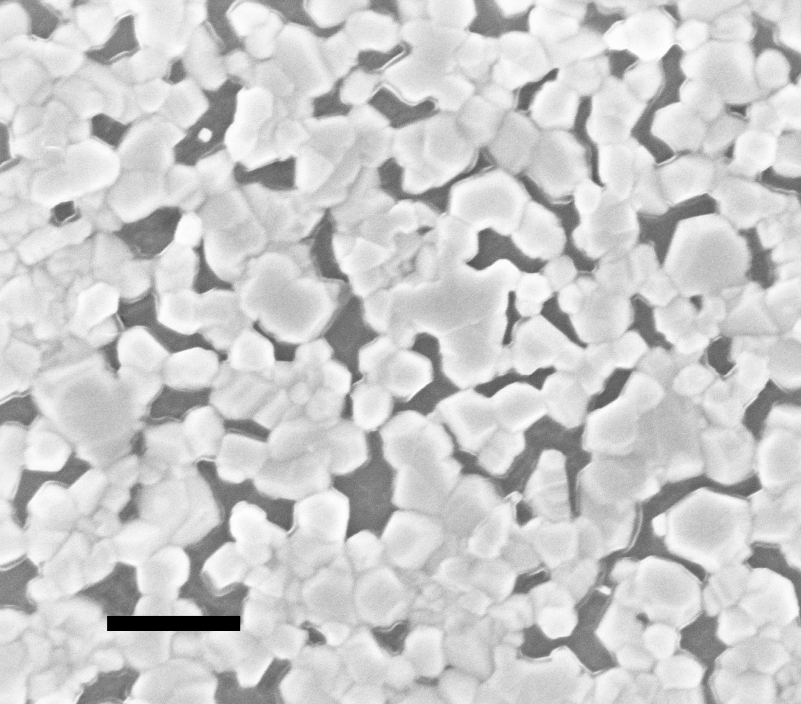}}
\large\raisebox{4cm}{b)}~\includegraphics[height=0.6\columnwidth]{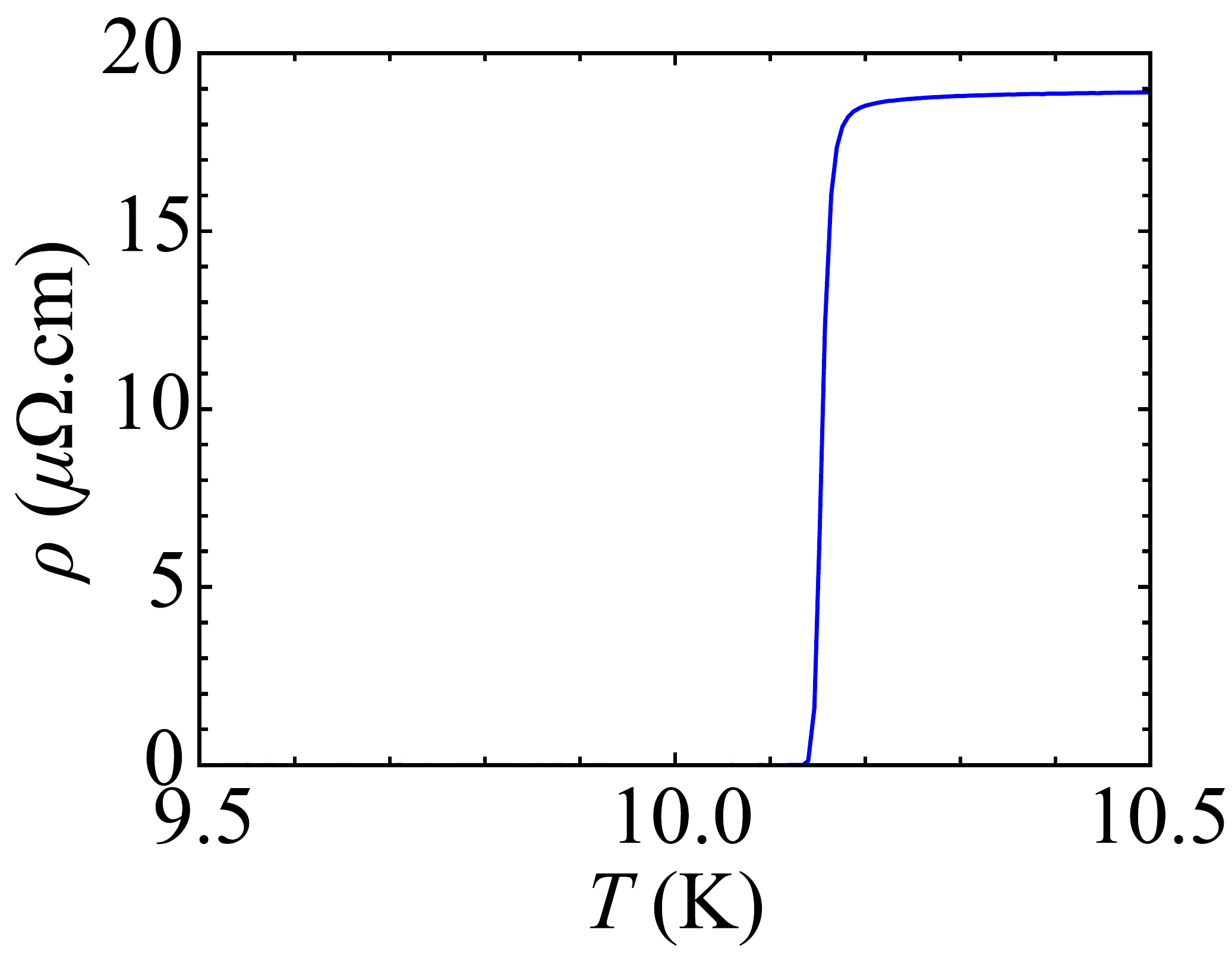}
\large\raisebox{4cm}{c)}~\includegraphics[height=0.55\columnwidth]{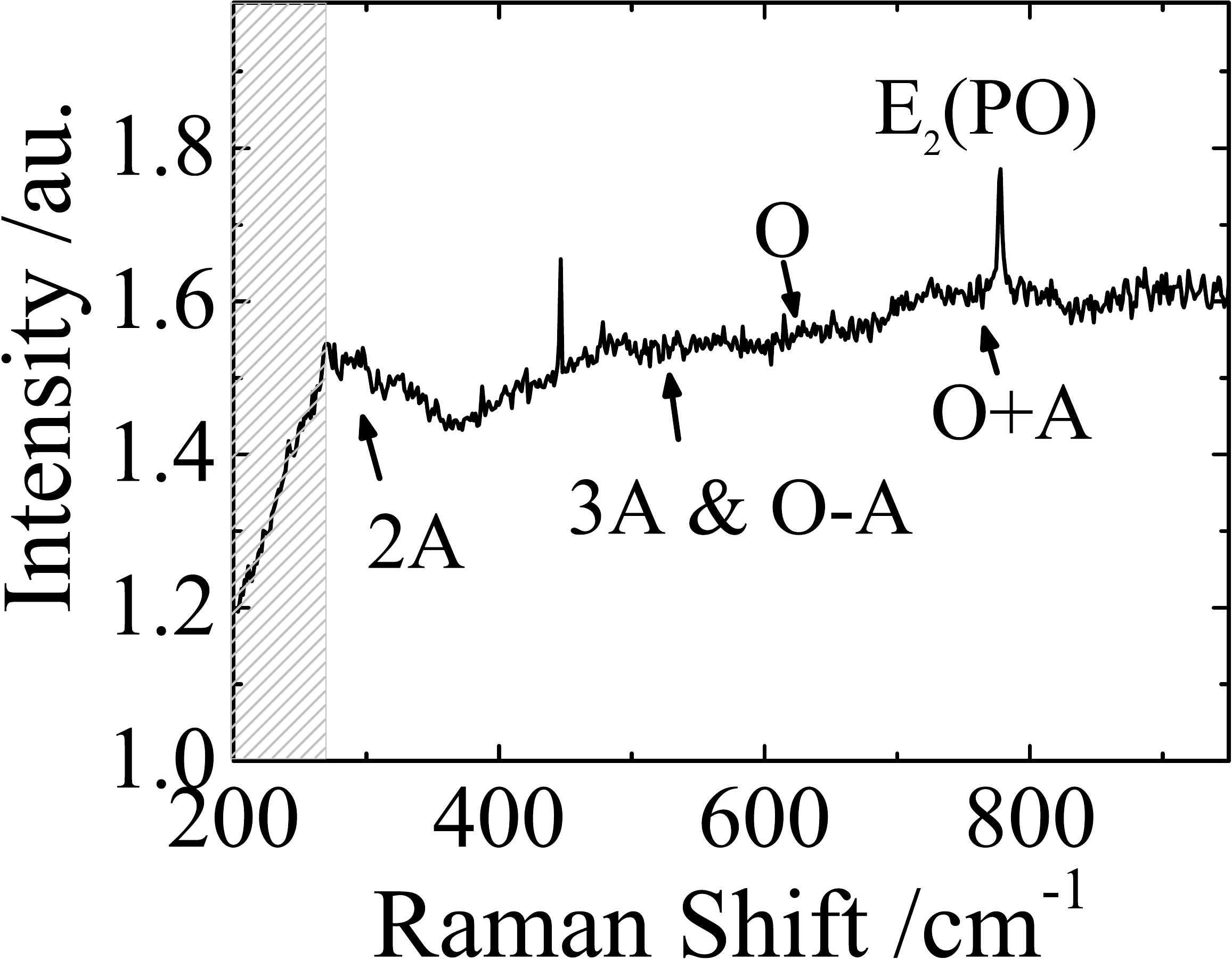}
\large\raisebox{4cm}{d)}~\includegraphics[height=0.55\columnwidth]{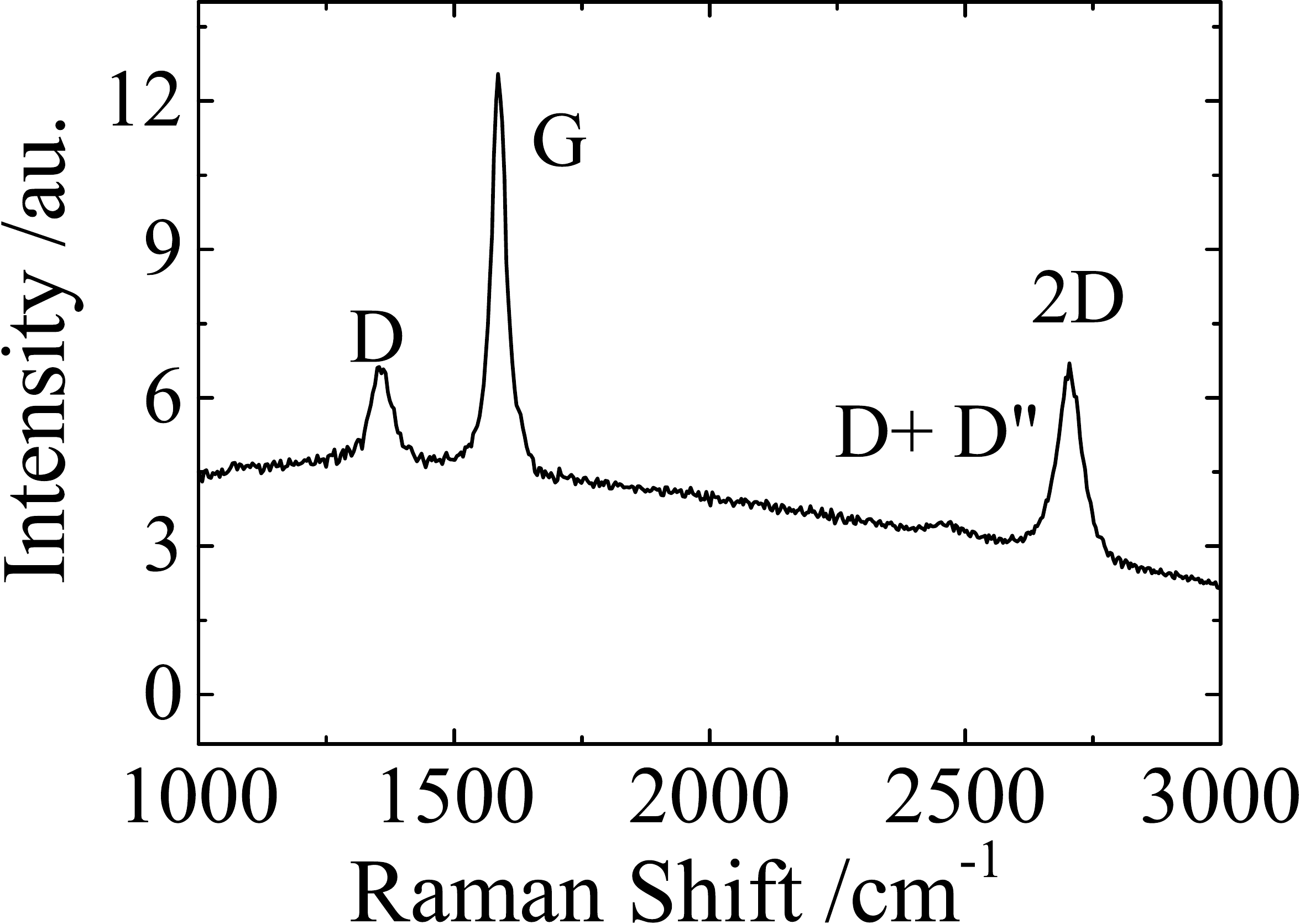}
\large\raisebox{4cm}{e)}~\includegraphics[height=0.55\columnwidth]{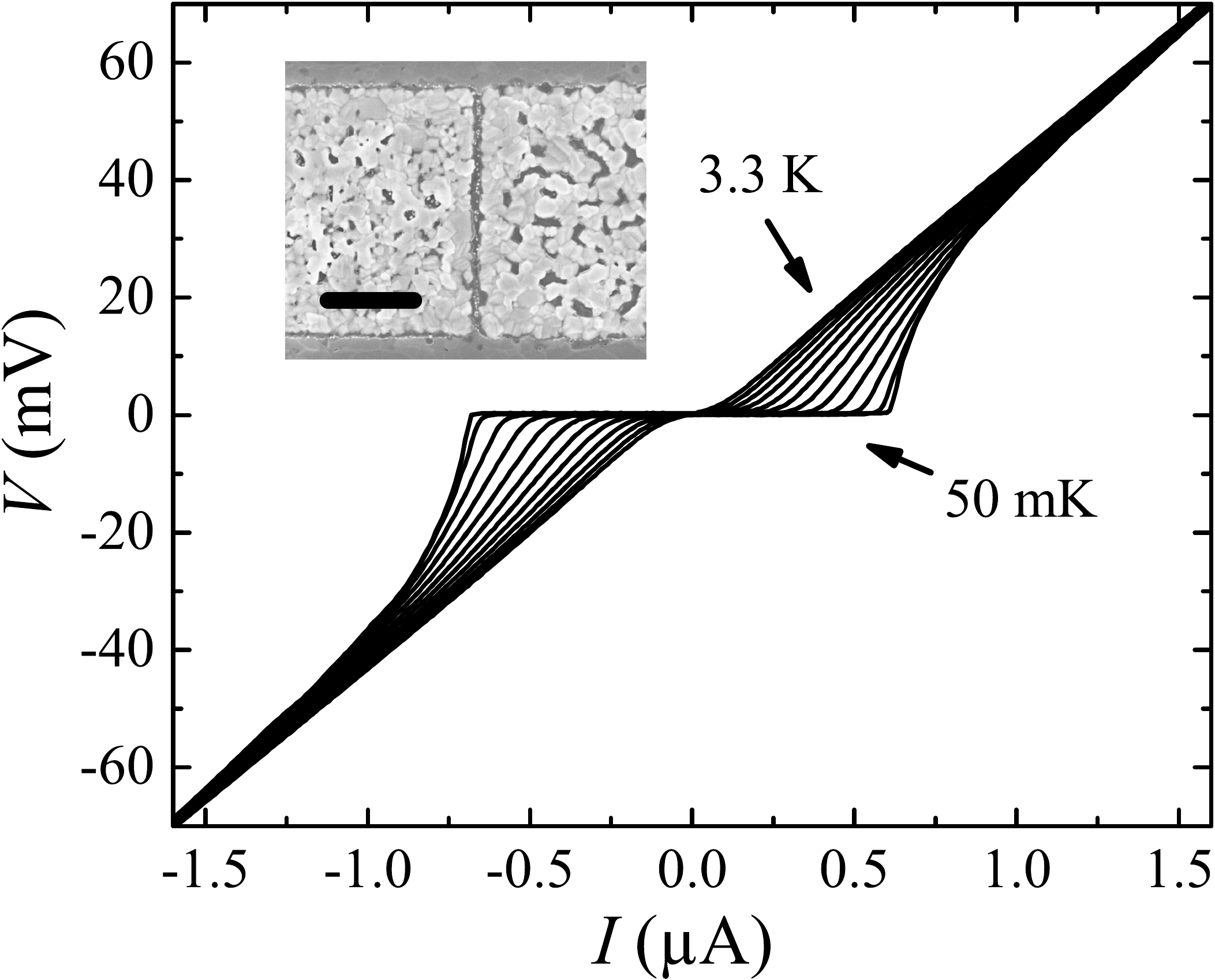}
\large\raisebox{4cm}{f)}~\includegraphics[height=0.55\columnwidth]{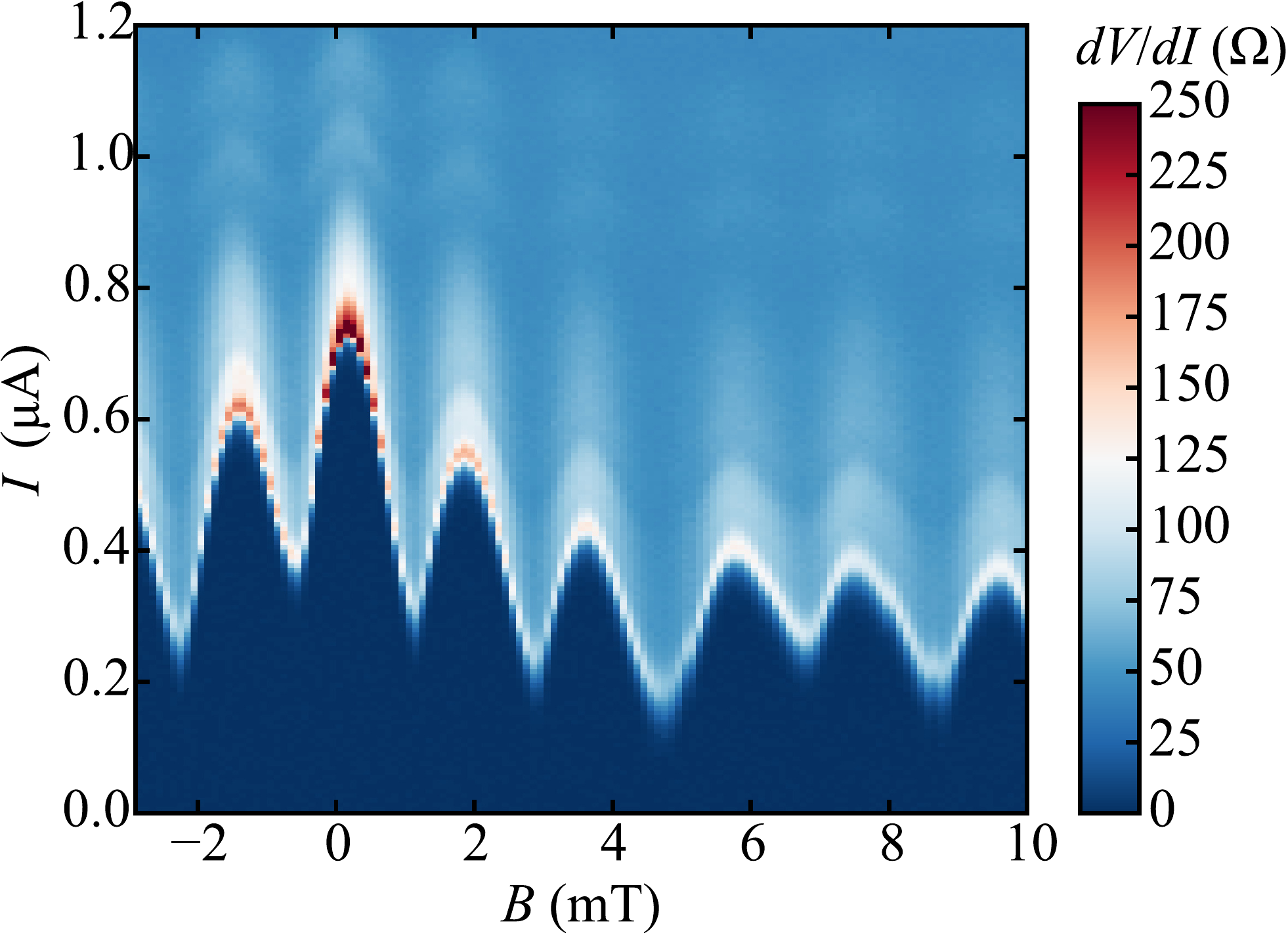}
\caption{a) SEM image of a SiC sample covered with an initially 20 nm thick layer of tantalum after annealing at 1360 $^\circ C$. The scale bar is 400 nm b) Resistance of the TaC film as function of temperature showing a superconducting transition at 10.15 K.  c) Low wavenumber Raman spectra acquired on the TaC sample. The shaded area shows the wave numbers where the signal is removed by the spectrometer's filter. d) Raman spectrum obtained on the same sample showing the typical graphene signal (G, 2D and D peaks) e)Temperature dependence of the current-voltage characteristic of a 70 nm junction (T=50 mK, and from 300 mK to 3.3 K in steps of 300 mK). The inset presents a SEM image of the junction (the scale bar is 800 nm). f) Differential resistance of the junction as function of current and magnetic field at 50 mK showing a Fraunhofer pattern.\label{Fig:TaC}}
\end{figure*}

\subsection{Josephson junctions}

We investigated the new possible functionalities offered by this technology by realizing Josephson junctions exploiting the superconducting properties of NbC. The junctions were prepared in the same way as described in Figure~\ref{Fig2} though here, electron beam lithography was used to achieve shorter channels length. For simplicity, after growth, the FGL were not etched in the form of ribbons between the NbC electrodes. The inset of Fig.~\ref{Fig3}a shows a SEM image of the junctions for which transport measurements are reported in the main panel. The transport data reported in Figure~\ref{Fig3}a reveal that a supercurrent flows in this junction below $T\approx1.5~$K when the Josephson energy overcomes thermal fluctuations ($\hbar I_c(T)/2e >k_BT $, with $I_c$ the critical current). This further confirms the good transparency of the electrical contacts since Josephson supercurrent is very sensitive to contact quality.

The magnetic field dependence of the critical current $I_c(H)$ shown in Figure~\ref{Fig3}b reveals the quantum interference effect between different current paths along the width of the junction.\cite{Tinkham1996} The observed dependence $I_c(H)$ shows deviations from the usual Fraunhofer pattern due to the peculiar geometry of the junction and subsequent non uniform current distribution.\cite{Dynes1971,BaronePaterno1982,Hart2014,Allen2015} Detailed discussions of the transport measurements can be found in supporting discussion 2.

\subsection{Generalization to another metallic carbide}
Finally, similar growth and device fabrication can be performed with Tantalum, demonstrating that the method can be generalized to other carbide forming metals. Figure~\ref{Fig:TaC}a shows a SEM image of a 20 nm thick ebeam evaporated tantalum (Ta) layer after annealing at  1360$^\circ$C. The morphology is similar to that of a niobium layer after the same treatment (Fig.~\ref{Fig1}a) with crystallization suggesting reaction with the substrate. We attribute the dewetting of the layer to the thiner starting layer compared to experiments with NbC. The solid state reaction to form TaC is confirmed by the superconducting temperature of the layer which is 10.15 K, i.e much higher than that of pure tantalum (4.45 K see Ref.~\cite{Milne1961}). Similarly to NbC the superconducting transition temperature of TaC is strongly dependent of its stoichiometry and 10.15 K allows to set a lower limit of 0.99 to the carbon/tantalum ratio. \cite{Giorgi1962} The good stoichiometry is confirmed by Raman measurements presented in Fig.~\ref{Fig:TaC}c where the Raman active optical and acoustic modes and their overtones suggest a low carbon vacancy concentration\cite{wipf_vacancy-induced_1981}. Furthermore the Raman measurements indicate that the Ta layer is covered with a graphitic layer as the characteristic G and 2D bands are observed. This set of observations suggests that tantalum follows the exact same route as Nb in reacting with SiC and forming few graphene layers on top.

We verified that the technique previously applied to NbC to fabricate Josephson junctions can be used to fabricate junctions with TaC contacts; Figure~\ref{Fig:TaC}e shows the current-voltage characteristics of such a junction (SEM image in the inset of Fig.~\ref{Fig:TaC}) and Fig.~\ref{Fig:TaC}f the magnetic field dependence. The junction is narrower and the superconducting current therefore set in at higher temperatures. Detailed discussions of the transport measurements can be found in supporting discussion 3.  

\section{Conclusion}
We have demonstrated the formation in a single annealing step of few graphene layers on SiC and their electrical contacts via solid state reactions between a carbide forming metal and the growth of graphene on the metallic carbide and on SiC. This original growth mechanism was exploited to fabricate FGL devices with low electrical contact resistance where the Josephson effect was observed. This contact preparation could be combined with selective graphene ribbon growth on SiC\cite{Camara2008,Sprinkle2010} to achieve single step circuit fabrication.

\section*{Acknowledgement}

TLQ was supported by a CIBLE fellowship from Region Rh\^{o}ne-Alpes. FLB acknowledges the support from Zeropova CEA program and Enhanced Eurotalents Marie Curie fellowship. We thank Val\'{e}rie Reita and the optics and microscopy technological group for valuable support on Raman spectroscopy, and Nedjma Bendiab for fruitful discussions. We thank C. Marcenat for lending his dilution fridge and technical assistance. We thank F. Gustavo for assistance during metal deposition.

\section*{Supplementary discussion}

Supplementary data related to this article can be found below.

\bibliography{Carbide}

\begin{thebibliography}{38}%
\makeatletter
\providecommand \@ifxundefined [1]{%
 \@ifx{#1\undefined}
}%
\providecommand \@ifnum [1]{%
 \ifnum #1\expandafter \@firstoftwo
 \else \expandafter \@secondoftwo
 \fi
}%
\providecommand \@ifx [1]{%
 \ifx #1\expandafter \@firstoftwo
 \else \expandafter \@secondoftwo
 \fi
}%
\providecommand \natexlab [1]{#1}%
\providecommand \enquote  [1]{``#1''}%
\providecommand \bibnamefont  [1]{#1}%
\providecommand \bibfnamefont [1]{#1}%
\providecommand \citenamefont [1]{#1}%
\providecommand \href@noop [0]{\@secondoftwo}%
\providecommand \href [0]{\begingroup \@sanitize@url \@href}%
\providecommand \@href[1]{\@@startlink{#1}\@@href}%
\providecommand \@@href[1]{\endgroup#1\@@endlink}%
\providecommand \@sanitize@url [0]{\catcode `\\12\catcode `\$12\catcode
  `\&12\catcode `\#12\catcode `\^12\catcode `\_12\catcode `\%12\relax}%
\providecommand \@@startlink[1]{}%
\providecommand \@@endlink[0]{}%
\providecommand \url  [0]{\begingroup\@sanitize@url \@url }%
\providecommand \@url [1]{\endgroup\@href {#1}{\urlprefix }}%
\providecommand \urlprefix  [0]{URL }%
\providecommand \Eprint [0]{\href }%
\providecommand \doibase [0]{http://dx.doi.org/}%
\providecommand \selectlanguage [0]{\@gobble}%
\providecommand \bibinfo  [0]{\@secondoftwo}%
\providecommand \bibfield  [0]{\@secondoftwo}%
\providecommand \translation [1]{[#1]}%
\providecommand \BibitemOpen [0]{}%
\providecommand \bibitemStop [0]{}%
\providecommand \bibitemNoStop [0]{.\EOS\space}%
\providecommand \EOS [0]{\spacefactor3000\relax}%
\providecommand \BibitemShut  [1]{\csname bibitem#1\endcsname}%
\let\auto@bib@innerbib\@empty
\bibitem [{\citenamefont {L{\'{e}}onard}\ and\ \citenamefont
  {Talin}(2011)}]{Leonard2011}%
  \BibitemOpen
  \bibfield  {author} {\bibinfo {author} {\bibfnamefont {F.}~\bibnamefont
  {L{\'{e}}onard}}\ and\ \bibinfo {author} {\bibfnamefont {A.~A.}\ \bibnamefont
  {Talin}},\ }\href {\doibase 10.1038/nnano.2011.196} {\bibfield  {journal}
  {\bibinfo  {journal} {Nature Nanotechnology}\ }\textbf {\bibinfo {volume}
  {6}},\ \bibinfo {pages} {773} (\bibinfo {year} {2011})}\BibitemShut {NoStop}%
\bibitem [{\citenamefont {Xia}\ \emph {et~al.}(2011)\citenamefont {Xia},
  \citenamefont {Perebeinos}, \citenamefont {Lin}, \citenamefont {Wu},\ and\
  \citenamefont {Avouris}}]{Xia2011}%
  \BibitemOpen
  \bibfield  {author} {\bibinfo {author} {\bibfnamefont {F.}~\bibnamefont
  {Xia}}, \bibinfo {author} {\bibfnamefont {V.}~\bibnamefont {Perebeinos}},
  \bibinfo {author} {\bibfnamefont {Y.-m.}\ \bibnamefont {Lin}}, \bibinfo
  {author} {\bibfnamefont {Y.}~\bibnamefont {Wu}}, \ and\ \bibinfo {author}
  {\bibfnamefont {P.}~\bibnamefont {Avouris}},\ }\href {\doibase
  10.1038/nnano.2011.6} {\bibfield  {journal} {\bibinfo  {journal} {Nature
  nanotechnology}\ }\textbf {\bibinfo {volume} {6}},\ \bibinfo {pages} {179}
  (\bibinfo {year} {2011})}\BibitemShut {NoStop}%
\bibitem [{\citenamefont {Allain}\ \emph {et~al.}(2015)\citenamefont {Allain},
  \citenamefont {Kang}, \citenamefont {Banerjee},\ and\ \citenamefont
  {Kis}}]{Allain2015}%
  \BibitemOpen
  \bibfield  {author} {\bibinfo {author} {\bibfnamefont {A.}~\bibnamefont
  {Allain}}, \bibinfo {author} {\bibfnamefont {J.}~\bibnamefont {Kang}},
  \bibinfo {author} {\bibfnamefont {K.}~\bibnamefont {Banerjee}}, \ and\
  \bibinfo {author} {\bibfnamefont {A.}~\bibnamefont {Kis}},\ }\href {\doibase
  10.1038/nmat4452} {\bibfield  {journal} {\bibinfo  {journal} {Nature
  Materials}\ }\textbf {\bibinfo {volume} {14}},\ \bibinfo {pages} {1195}
  (\bibinfo {year} {2015})}\BibitemShut {NoStop}%
\bibitem [{\citenamefont {Robinson}\ \emph {et~al.}(2011)\citenamefont
  {Robinson}, \citenamefont {LaBella}, \citenamefont {Zhu}, \citenamefont
  {Hollander}, \citenamefont {Kasarda}, \citenamefont {Hughes}, \citenamefont
  {Trumbull}, \citenamefont {Cavalero},\ and\ \citenamefont
  {Snyder}}]{Robinson2011}%
  \BibitemOpen
  \bibfield  {author} {\bibinfo {author} {\bibfnamefont {J.~a.}\ \bibnamefont
  {Robinson}}, \bibinfo {author} {\bibfnamefont {M.}~\bibnamefont {LaBella}},
  \bibinfo {author} {\bibfnamefont {M.}~\bibnamefont {Zhu}}, \bibinfo {author}
  {\bibfnamefont {M.}~\bibnamefont {Hollander}}, \bibinfo {author}
  {\bibfnamefont {R.}~\bibnamefont {Kasarda}}, \bibinfo {author} {\bibfnamefont
  {Z.}~\bibnamefont {Hughes}}, \bibinfo {author} {\bibfnamefont
  {K.}~\bibnamefont {Trumbull}}, \bibinfo {author} {\bibfnamefont
  {R.}~\bibnamefont {Cavalero}}, \ and\ \bibinfo {author} {\bibfnamefont
  {D.}~\bibnamefont {Snyder}},\ }\href {\doibase 10.1063/1.3549183} {\bibfield
  {journal} {\bibinfo  {journal} {Applied Physics Letters}\ }\textbf {\bibinfo
  {volume} {98}},\ \bibinfo {pages} {053103} (\bibinfo {year}
  {2011})}\BibitemShut {NoStop}%
\bibitem [{\citenamefont {Wang}\ \emph {et~al.}(2013)\citenamefont {Wang},
  \citenamefont {Meric}, \citenamefont {Huang}, \citenamefont {Gao},
  \citenamefont {Gao}, \citenamefont {Tran}, \citenamefont {Taniguchi},
  \citenamefont {Watanabe}, \citenamefont {Campos}, \citenamefont {Muller},
  \citenamefont {Guo}, \citenamefont {Kim}, \citenamefont {Hone}, \citenamefont
  {Shepard},\ and\ \citenamefont {Dean}}]{Wang2013}%
  \BibitemOpen
  \bibfield  {author} {\bibinfo {author} {\bibfnamefont {L.}~\bibnamefont
  {Wang}}, \bibinfo {author} {\bibfnamefont {I.}~\bibnamefont {Meric}},
  \bibinfo {author} {\bibfnamefont {P.~Y.}\ \bibnamefont {Huang}}, \bibinfo
  {author} {\bibfnamefont {Q.}~\bibnamefont {Gao}}, \bibinfo {author}
  {\bibfnamefont {Y.}~\bibnamefont {Gao}}, \bibinfo {author} {\bibfnamefont
  {H.}~\bibnamefont {Tran}}, \bibinfo {author} {\bibfnamefont {T.}~\bibnamefont
  {Taniguchi}}, \bibinfo {author} {\bibfnamefont {K.}~\bibnamefont {Watanabe}},
  \bibinfo {author} {\bibfnamefont {L.~M.}\ \bibnamefont {Campos}}, \bibinfo
  {author} {\bibfnamefont {D.~a.}\ \bibnamefont {Muller}}, \bibinfo {author}
  {\bibfnamefont {J.}~\bibnamefont {Guo}}, \bibinfo {author} {\bibfnamefont
  {P.}~\bibnamefont {Kim}}, \bibinfo {author} {\bibfnamefont {J.}~\bibnamefont
  {Hone}}, \bibinfo {author} {\bibfnamefont {K.~L.}\ \bibnamefont {Shepard}}, \
  and\ \bibinfo {author} {\bibfnamefont {C.~R.}\ \bibnamefont {Dean}},\ }\href
  {\doibase 10.1126/science.1244358} {\bibfield  {journal} {\bibinfo  {journal}
  {Science (New York, N.Y.)}\ }\textbf {\bibinfo {volume} {342}},\ \bibinfo
  {pages} {614} (\bibinfo {year} {2013})}\BibitemShut {NoStop}%
\bibitem [{\citenamefont {Park}\ \emph {et~al.}(2016)\citenamefont {Park},
  \citenamefont {Jung}, \citenamefont {Kang}, \citenamefont {Jeon},
  \citenamefont {Yoo}, \citenamefont {Park}, \citenamefont {Lee}, \citenamefont
  {Jang}, \citenamefont {Lee}, \citenamefont {Park}, \citenamefont {Yu},
  \citenamefont {Shin}, \citenamefont {Lee},\ and\ \citenamefont
  {Park}}]{Park2016}%
  \BibitemOpen
  \bibfield  {author} {\bibinfo {author} {\bibfnamefont {H.-y.}\ \bibnamefont
  {Park}}, \bibinfo {author} {\bibfnamefont {W.-s.}\ \bibnamefont {Jung}},
  \bibinfo {author} {\bibfnamefont {D.-h.}\ \bibnamefont {Kang}}, \bibinfo
  {author} {\bibfnamefont {J.}~\bibnamefont {Jeon}}, \bibinfo {author}
  {\bibfnamefont {G.}~\bibnamefont {Yoo}}, \bibinfo {author} {\bibfnamefont
  {Y.}~\bibnamefont {Park}}, \bibinfo {author} {\bibfnamefont {J.}~\bibnamefont
  {Lee}}, \bibinfo {author} {\bibfnamefont {Y.~H.}\ \bibnamefont {Jang}},
  \bibinfo {author} {\bibfnamefont {J.}~\bibnamefont {Lee}}, \bibinfo {author}
  {\bibfnamefont {S.}~\bibnamefont {Park}}, \bibinfo {author} {\bibfnamefont
  {H.-y.}\ \bibnamefont {Yu}}, \bibinfo {author} {\bibfnamefont
  {B.}~\bibnamefont {Shin}}, \bibinfo {author} {\bibfnamefont {S.}~\bibnamefont
  {Lee}}, \ and\ \bibinfo {author} {\bibfnamefont {J.-h.}\ \bibnamefont
  {Park}},\ }\href {\doibase 10.1002/adma.201503715} {\ ,\ \bibinfo {pages}
  {864} (\bibinfo {year} {2016})}\BibitemShut {NoStop}%
\bibitem [{\citenamefont {Borovikov}\ and\ \citenamefont
  {Zangwill}(2009)}]{Borovikov2009}%
  \BibitemOpen
  \bibfield  {author} {\bibinfo {author} {\bibfnamefont {V.}~\bibnamefont
  {Borovikov}}\ and\ \bibinfo {author} {\bibfnamefont {A.}~\bibnamefont
  {Zangwill}},\ }\href {\doibase 10.1103/PhysRevB.80.121406} {\bibfield
  {journal} {\bibinfo  {journal} {Phys. Rev. B}\ }\textbf {\bibinfo {volume}
  {80}},\ \bibinfo {pages} {121406} (\bibinfo {year} {2009})}\BibitemShut
  {NoStop}%
\bibitem [{\citenamefont {Ohta}\ \emph {et~al.}(2010)\citenamefont {Ohta},
  \citenamefont {Bartelt}, \citenamefont {Nie}, \citenamefont {Th\"urmer},\
  and\ \citenamefont {Kellogg}}]{Taisuke2010}%
  \BibitemOpen
  \bibfield  {author} {\bibinfo {author} {\bibfnamefont {T.}~\bibnamefont
  {Ohta}}, \bibinfo {author} {\bibfnamefont {N.~C.}\ \bibnamefont {Bartelt}},
  \bibinfo {author} {\bibfnamefont {S.}~\bibnamefont {Nie}}, \bibinfo {author}
  {\bibfnamefont {K.}~\bibnamefont {Th\"urmer}}, \ and\ \bibinfo {author}
  {\bibfnamefont {G.~L.}\ \bibnamefont {Kellogg}},\ }\href {\doibase
  10.1103/PhysRevB.81.121411} {\bibfield  {journal} {\bibinfo  {journal} {Phys.
  Rev. B}\ }\textbf {\bibinfo {volume} {81}},\ \bibinfo {pages} {121411}
  (\bibinfo {year} {2010})}\BibitemShut {NoStop}%
\bibitem [{\citenamefont {Kusunoki}\ \emph {et~al.}(2015)\citenamefont
  {Kusunoki}, \citenamefont {Norimatsu}, \citenamefont {Bao}, \citenamefont
  {Morita},\ and\ \citenamefont {Starke}}]{Kusunoki2015}%
  \BibitemOpen
  \bibfield  {author} {\bibinfo {author} {\bibfnamefont {M.}~\bibnamefont
  {Kusunoki}}, \bibinfo {author} {\bibfnamefont {W.}~\bibnamefont {Norimatsu}},
  \bibinfo {author} {\bibfnamefont {J.}~\bibnamefont {Bao}}, \bibinfo {author}
  {\bibfnamefont {K.}~\bibnamefont {Morita}}, \ and\ \bibinfo {author}
  {\bibfnamefont {U.}~\bibnamefont {Starke}},\ }\href {\doibase
  10.7566/JPSJ.84.121014} {\bibfield  {journal} {\bibinfo  {journal} {Journal
  of the Physical Society of Japan}\ }\textbf {\bibinfo {volume} {84}},\
  \bibinfo {pages} {121014} (\bibinfo {year} {2015})}\BibitemShut {NoStop}%
\bibitem [{\citenamefont {Berger}\ \emph {et~al.}(2004)\citenamefont {Berger},
  \citenamefont {Song}, \citenamefont {Li}, \citenamefont {Li}, \citenamefont
  {Ogbarzghi}, \citenamefont {Feng},\ and\ \citenamefont {\textit{et
  al.}}}]{Berger2004}%
  \BibitemOpen
  \bibfield  {author} {\bibinfo {author} {\bibfnamefont {C.}~\bibnamefont
  {Berger}}, \bibinfo {author} {\bibfnamefont {Z.}~\bibnamefont {Song}},
  \bibinfo {author} {\bibfnamefont {T.}~\bibnamefont {Li}}, \bibinfo {author}
  {\bibfnamefont {A.}~\bibnamefont {Li}}, \bibinfo {author} {\bibfnamefont
  {A.~Y.}\ \bibnamefont {Ogbarzghi}}, \bibinfo {author} {\bibfnamefont
  {R.}~\bibnamefont {Feng}}, \ and\ \bibinfo {author} {\bibnamefont {\textit{et
  al.}}},\ }\href {\doibase 10.1021/jp040650f} {\bibfield  {journal} {\bibinfo
  {journal} {J. Phys. Chem B}\ }\textbf {\bibinfo {volume} {108}},\ \bibinfo
  {pages} {19912} (\bibinfo {year} {2004})}\BibitemShut {NoStop}%
\bibitem [{\citenamefont {Berger}\ \emph {et~al.}(2006)\citenamefont {Berger},
  \citenamefont {Song}, \citenamefont {Li}, \citenamefont {Wu}, \citenamefont
  {Brown}, \citenamefont {Naud}, \citenamefont {Mayou}, \citenamefont {Li},
  \citenamefont {Hass}, \citenamefont {Marchenkov},\ and\ \citenamefont
  {Others}}]{Berger2006}%
  \BibitemOpen
  \bibfield  {author} {\bibinfo {author} {\bibfnamefont {C.}~\bibnamefont
  {Berger}}, \bibinfo {author} {\bibfnamefont {Z.}~\bibnamefont {Song}},
  \bibinfo {author} {\bibfnamefont {X.}~\bibnamefont {Li}}, \bibinfo {author}
  {\bibfnamefont {X.}~\bibnamefont {Wu}}, \bibinfo {author} {\bibfnamefont
  {N.}~\bibnamefont {Brown}}, \bibinfo {author} {\bibfnamefont
  {C.}~\bibnamefont {Naud}}, \bibinfo {author} {\bibfnamefont {D.}~\bibnamefont
  {Mayou}}, \bibinfo {author} {\bibfnamefont {T.}~\bibnamefont {Li}}, \bibinfo
  {author} {\bibfnamefont {J.}~\bibnamefont {Hass}}, \bibinfo {author}
  {\bibfnamefont {A.~N.}\ \bibnamefont {Marchenkov}}, \ and\ \bibinfo {author}
  {\bibnamefont {Others}},\ }\href@noop {} {\bibfield  {journal} {\bibinfo
  {journal} {Science}\ }\textbf {\bibinfo {volume} {312}},\ \bibinfo {pages}
  {1191} (\bibinfo {year} {2006})}\BibitemShut {NoStop}%
\bibitem [{\citenamefont {Foster}\ \emph {et~al.}(1958)\citenamefont {Foster},
  \citenamefont {Long},\ and\ \citenamefont {C.}}]{Foster1958}%
  \BibitemOpen
  \bibfield  {author} {\bibinfo {author} {\bibfnamefont {L.~M.}\ \bibnamefont
  {Foster}}, \bibinfo {author} {\bibfnamefont {G.}~\bibnamefont {Long}}, \ and\
  \bibinfo {author} {\bibfnamefont {S.~H.}\ \bibnamefont {C.}},\ }\href@noop {}
  {\bibfield  {journal} {\bibinfo  {journal} {Amer. Miner.}\ }\textbf {\bibinfo
  {volume} {43}},\ \bibinfo {pages} {285} (\bibinfo {year} {1958})}\BibitemShut
  {NoStop}%
\bibitem [{\citenamefont {Presser}\ \emph {et~al.}(2011)\citenamefont
  {Presser}, \citenamefont {Heon},\ and\ \citenamefont
  {Gogotsi}}]{Presser2011}%
  \BibitemOpen
  \bibfield  {author} {\bibinfo {author} {\bibfnamefont {V.}~\bibnamefont
  {Presser}}, \bibinfo {author} {\bibfnamefont {M.}~\bibnamefont {Heon}}, \
  and\ \bibinfo {author} {\bibfnamefont {Y.}~\bibnamefont {Gogotsi}},\ }\href
  {\doibase 10.1002/adfm.201002094} {\bibfield  {journal} {\bibinfo  {journal}
  {Advanced Functional Materials}\ }\textbf {\bibinfo {volume} {21}},\ \bibinfo
  {pages} {810} (\bibinfo {year} {2011})}\BibitemShut {NoStop}%
\bibitem [{\citenamefont {Aizawa}\ \emph {et~al.}(1992)\citenamefont {Aizawa},
  \citenamefont {Hwang}, \citenamefont {Hayami}, \citenamefont {Souda},
  \citenamefont {Otani},\ and\ \citenamefont {Ishizawa}}]{Aizawa1992}%
  \BibitemOpen
  \bibfield  {author} {\bibinfo {author} {\bibfnamefont {T.}~\bibnamefont
  {Aizawa}}, \bibinfo {author} {\bibfnamefont {Y.}~\bibnamefont {Hwang}},
  \bibinfo {author} {\bibfnamefont {W.}~\bibnamefont {Hayami}}, \bibinfo
  {author} {\bibfnamefont {R.}~\bibnamefont {Souda}}, \bibinfo {author}
  {\bibfnamefont {S.}~\bibnamefont {Otani}}, \ and\ \bibinfo {author}
  {\bibfnamefont {Y.}~\bibnamefont {Ishizawa}},\ }\href {\doibase
  10.1016/0039-6028(92)90046-9} {\bibfield  {journal} {\bibinfo  {journal}
  {Surface Science}\ }\textbf {\bibinfo {volume} {260}},\ \bibinfo {pages}
  {311} (\bibinfo {year} {1992})}\BibitemShut {NoStop}%
\bibitem [{\citenamefont {Hwang}\ \emph {et~al.}(1992)\citenamefont {Hwang},
  \citenamefont {Aizawa}, \citenamefont {Hayami}, \citenamefont {Otani},
  \citenamefont {Ishizawa},\ and\ \citenamefont {Park}}]{Hwang1992}%
  \BibitemOpen
  \bibfield  {author} {\bibinfo {author} {\bibfnamefont {T.}~\bibnamefont
  {Hwang}}, \bibinfo {author} {\bibfnamefont {W.}~\bibnamefont {Aizawa}},
  \bibinfo {author} {\bibfnamefont {S.}~\bibnamefont {Hayami}}, \bibinfo
  {author} {\bibfnamefont {S.}~\bibnamefont {Otani}}, \bibinfo {author}
  {\bibfnamefont {Y.}~\bibnamefont {Ishizawa}}, \ and\ \bibinfo {author}
  {\bibfnamefont {S.~J.}\ \bibnamefont {Park}},\ }\href {\doibase
  10.1016/0038-1098(92)90765-2} {\bibfield  {journal} {\bibinfo  {journal}
  {Solid State Communications}\ }\textbf {\bibinfo {volume} {81}},\ \bibinfo
  {pages} {397} (\bibinfo {year} {1992})}\BibitemShut {NoStop}%
\bibitem [{\citenamefont {Burykina}\ \emph {et~al.}(1968)\citenamefont
  {Burykina}, \citenamefont {Strashinskaya},\ and\ \citenamefont
  {Evtushok}}]{Burykina1968}%
  \BibitemOpen
  \bibfield  {author} {\bibinfo {author} {\bibfnamefont {A.~L.}\ \bibnamefont
  {Burykina}}, \bibinfo {author} {\bibfnamefont {L.~V.}\ \bibnamefont
  {Strashinskaya}}, \ and\ \bibinfo {author} {\bibfnamefont {T.~M.}\
  \bibnamefont {Evtushok}},\ }\href {\doibase 10.1007/BF00938092} {\bibfield
  {journal} {\bibinfo  {journal} {Fiziko-Kimicheskaya Mekhanika Materialov}\
  }\textbf {\bibinfo {volume} {4}},\ \bibinfo {pages} {301} (\bibinfo {year}
  {1968})}\BibitemShut {NoStop}%
\bibitem [{\citenamefont {Yaney}\ and\ \citenamefont
  {Joshi}(1990)}]{Yaney1990}%
  \BibitemOpen
  \bibfield  {author} {\bibinfo {author} {\bibfnamefont {D.~L.}\ \bibnamefont
  {Yaney}}\ and\ \bibinfo {author} {\bibfnamefont {A.}~\bibnamefont {Joshi}},\
  }\href {\doibase 10.1557/JMR.1990.2197} {\bibfield  {journal} {\bibinfo
  {journal} {J. Mater. Res.}\ }\textbf {\bibinfo {volume} {5}},\ \bibinfo
  {pages} {10} (\bibinfo {year} {1990})}\BibitemShut {NoStop}%
\bibitem [{\citenamefont {Chou}\ \emph {et~al.}(1990)\citenamefont {Chou},
  \citenamefont {Joshi},\ and\ \citenamefont {Waldsworth}}]{Chou1990}%
  \BibitemOpen
  \bibfield  {author} {\bibinfo {author} {\bibfnamefont {T.~C.}\ \bibnamefont
  {Chou}}, \bibinfo {author} {\bibfnamefont {A.}~\bibnamefont {Joshi}}, \ and\
  \bibinfo {author} {\bibfnamefont {J.}~\bibnamefont {Waldsworth}},\ }\href
  {\doibase 10.1116/1.577673} {\bibfield  {journal} {\bibinfo  {journal}
  {Journal of Vacuum Science Technology A}\ }\textbf {\bibinfo {volume} {9}},\
  \bibinfo {pages} {1525} (\bibinfo {year} {1990})}\BibitemShut {NoStop}%
\bibitem [{\citenamefont {Wang}\ \emph {et~al.}(2009)\citenamefont {Wang},
  \citenamefont {Liu}, \citenamefont {Zhang},\ and\ \citenamefont
  {Cheng}}]{Wang2009}%
  \BibitemOpen
  \bibfield  {author} {\bibinfo {author} {\bibfnamefont {Y.}~\bibnamefont
  {Wang}}, \bibinfo {author} {\bibfnamefont {Q.}~\bibnamefont {Liu}}, \bibinfo
  {author} {\bibfnamefont {L.}~\bibnamefont {Zhang}}, \ and\ \bibinfo {author}
  {\bibfnamefont {L.}~\bibnamefont {Cheng}},\ }\href {\doibase
  10.1007/s11998-008-9129-1} {\bibfield  {journal} {\bibinfo  {journal}
  {Journal of coatings technology research}\ }\textbf {\bibinfo {volume} {6}},\
  \bibinfo {pages} {413} (\bibinfo {year} {2009})}\BibitemShut {NoStop}%
\bibitem [{\citenamefont {Van~der Pauw}(1958)}]{Vanderpauw1958}%
  \BibitemOpen
  \bibfield  {author} {\bibinfo {author} {\bibfnamefont {L.}~\bibnamefont
  {Van~der Pauw}},\ }\href@noop {} {\bibfield  {journal} {\bibinfo  {journal}
  {Philips Research Reports}\ }\textbf {\bibinfo {volume} {13}},\ \bibinfo
  {pages} {1} (\bibinfo {year} {1958})}\BibitemShut {NoStop}%
\bibitem [{\citenamefont {N.~R.~Werthamer}(1966)}]{Werthamer1966}%
  \BibitemOpen
  \bibfield  {author} {\bibinfo {author} {\bibfnamefont {P.~H.}\ \bibnamefont
  {N.~R.~Werthamer}, \bibfnamefont {E.~Helfand}},\ }\href {\doibase
  10.1103/PhysRev.147.295} {\bibfield  {journal} {\bibinfo  {journal} {Phys.
  Rev. B.}\ }\textbf {\bibinfo {volume} {147}},\ \bibinfo {pages} {295}
  (\bibinfo {year} {1966})}\BibitemShut {NoStop}%
\bibitem [{\citenamefont {Karasik}\ \emph {et~al.}(1996)\citenamefont
  {Karasik}, \citenamefont {IL'in}, \citenamefont {Pechen},\ and\ \citenamefont
  {Krasnosvobodtsev}}]{Karasik1996}%
  \BibitemOpen
  \bibfield  {author} {\bibinfo {author} {\bibfnamefont {B.~S.}\ \bibnamefont
  {Karasik}}, \bibinfo {author} {\bibfnamefont {K.~S.}\ \bibnamefont {IL'in}},
  \bibinfo {author} {\bibfnamefont {E.~V.}\ \bibnamefont {Pechen}}, \ and\
  \bibinfo {author} {\bibfnamefont {S.~I.}\ \bibnamefont {Krasnosvobodtsev}},\
  }\href {\doibase 10.1063/1.115886} {\bibfield  {journal} {\bibinfo  {journal}
  {Appl. Phys. Lett}\ }\textbf {\bibinfo {volume} {68}},\ \bibinfo {pages}
  {2285} (\bibinfo {year} {1996})}\BibitemShut {NoStop}%
\bibitem [{\citenamefont {Golovashkin}\ \emph {et~al.}(1986)\citenamefont
  {Golovashkin}, \citenamefont {Zhurkin}, \citenamefont {Karuzskii},
  \citenamefont {Krasnosvobodtsev}, \citenamefont {Martoviskii}, \citenamefont
  {Pechen'},\ and\ \citenamefont {\textit{et al.}}}]{Golovashkin1986}%
  \BibitemOpen
  \bibfield  {author} {\bibinfo {author} {\bibfnamefont {A.~I.}\ \bibnamefont
  {Golovashkin}}, \bibinfo {author} {\bibfnamefont {B.~G.}\ \bibnamefont
  {Zhurkin}}, \bibinfo {author} {\bibfnamefont {A.~L.}\ \bibnamefont
  {Karuzskii}}, \bibinfo {author} {\bibfnamefont {S.~I.}\ \bibnamefont
  {Krasnosvobodtsev}}, \bibinfo {author} {\bibfnamefont {V.~P.}\ \bibnamefont
  {Martoviskii}}, \bibinfo {author} {\bibfnamefont {E.~V.}\ \bibnamefont
  {Pechen'}}, \ and\ \bibinfo {author} {\bibnamefont {\textit{et al.}}},\
  }\href@noop {} {\bibfield  {journal} {\bibinfo  {journal} {Sov. Phys. Solid
  State}\ }\textbf {\bibinfo {volume} {28}},\ \bibinfo {pages} {1881} (\bibinfo
  {year} {1986})}\BibitemShut {NoStop}%
\bibitem [{\citenamefont {Kimura}\ \emph {et~al.}(2013)\citenamefont {Kimura},
  \citenamefont {Shoji}, \citenamefont {Yamamoto}, \citenamefont {Norimatsu},\
  and\ \citenamefont {Kusunoki}}]{Kimura2013}%
  \BibitemOpen
  \bibfield  {author} {\bibinfo {author} {\bibfnamefont {K.}~\bibnamefont
  {Kimura}}, \bibinfo {author} {\bibfnamefont {K.}~\bibnamefont {Shoji}},
  \bibinfo {author} {\bibfnamefont {Y.}~\bibnamefont {Yamamoto}}, \bibinfo
  {author} {\bibfnamefont {W.}~\bibnamefont {Norimatsu}}, \ and\ \bibinfo
  {author} {\bibfnamefont {M.}~\bibnamefont {Kusunoki}},\ }\href {\doibase
  10.1103/PhysRevB.87.075431} {\bibfield  {journal} {\bibinfo  {journal} {Phys.
  Rev. B}\ }\textbf {\bibinfo {volume} {87}},\ \bibinfo {pages} {075431}
  (\bibinfo {year} {2013})}\BibitemShut {NoStop}%
\bibitem [{\citenamefont {Wipf}\ \emph {et~al.}(1981)\citenamefont {Wipf},
  \citenamefont {Klein},\ and\ \citenamefont
  {Williams}}]{wipf_vacancy-induced_1981}%
  \BibitemOpen
  \bibfield  {author} {\bibinfo {author} {\bibfnamefont {H.}~\bibnamefont
  {Wipf}}, \bibinfo {author} {\bibfnamefont {M.}~\bibnamefont {Klein}}, \ and\
  \bibinfo {author} {\bibfnamefont {W.}~\bibnamefont {Williams}},\ }\href
  {\doibase 10.1002/pssb.2221080225} {\bibfield  {journal} {\bibinfo  {journal}
  {Phys. Stat. Sol. (b)}\ }\textbf {\bibinfo {volume} {108}},\ \bibinfo {pages}
  {489} (\bibinfo {year} {1981})}\BibitemShut {NoStop}%
\bibitem [{\citenamefont {Dubistky}\ \emph {et~al.}(2005)\citenamefont
  {Dubistky}, \citenamefont {Blank}, \citenamefont {Buga}, \citenamefont
  {Semenova}, \citenamefont {Kul'bachinski}, \citenamefont {Krechetov},\ and\
  \citenamefont {Kytin}}]{Dubistky2005}%
  \BibitemOpen
  \bibfield  {author} {\bibinfo {author} {\bibfnamefont {G.~A.}\ \bibnamefont
  {Dubistky}}, \bibinfo {author} {\bibfnamefont {V.~D.}\ \bibnamefont {Blank}},
  \bibinfo {author} {\bibfnamefont {S.~G.}\ \bibnamefont {Buga}}, \bibinfo
  {author} {\bibfnamefont {E.~E.}\ \bibnamefont {Semenova}}, \bibinfo {author}
  {\bibfnamefont {V.~A.}\ \bibnamefont {Kul'bachinski}}, \bibinfo {author}
  {\bibfnamefont {A.~V.}\ \bibnamefont {Krechetov}}, \ and\ \bibinfo {author}
  {\bibfnamefont {V.~G.}\ \bibnamefont {Kytin}},\ }\href {\doibase
  10.1134/1.1931011} {\bibfield  {journal} {\bibinfo  {journal} {JETP Lett.}\
  }\textbf {\bibinfo {volume} {81}},\ \bibinfo {pages} {260} (\bibinfo {year}
  {2005})}\BibitemShut {NoStop}%
\bibitem [{\citenamefont {Giorgi}\ \emph {et~al.}(1962)\citenamefont {Giorgi},
  \citenamefont {Szklarz}, \citenamefont {Stroms}, \citenamefont {Bowman},\
  and\ \citenamefont {Matthias}}]{Giorgi1962}%
  \BibitemOpen
  \bibfield  {author} {\bibinfo {author} {\bibfnamefont {A.~L.}\ \bibnamefont
  {Giorgi}}, \bibinfo {author} {\bibfnamefont {E.~G.}\ \bibnamefont {Szklarz}},
  \bibinfo {author} {\bibfnamefont {E.~K.}\ \bibnamefont {Stroms}}, \bibinfo
  {author} {\bibfnamefont {A.~L.}\ \bibnamefont {Bowman}}, \ and\ \bibinfo
  {author} {\bibfnamefont {B.~T.}\ \bibnamefont {Matthias}},\ }\href {\doibase
  10.1103/PhysRev.125.837} {\bibfield  {journal} {\bibinfo  {journal} {Phys.
  Rev.}\ }\textbf {\bibinfo {volume} {125}},\ \bibinfo {pages} {837} (\bibinfo
  {year} {1962})}\BibitemShut {NoStop}%
\bibitem [{\citenamefont {Berger}(1972)}]{Berger1972}%
  \BibitemOpen
  \bibfield  {author} {\bibinfo {author} {\bibfnamefont {H.~H.}\ \bibnamefont
  {Berger}},\ }\href {\doibase 10.1149/1.2404240} {\bibfield  {journal}
  {\bibinfo  {journal} {Journal of The Electrochemical Society}\ }\textbf
  {\bibinfo {volume} {119}},\ \bibinfo {pages} {507} (\bibinfo {year}
  {1972})}\BibitemShut {NoStop}%
\bibitem [{\citenamefont {Ito}\ \emph {et~al.}(2015)\citenamefont {Ito},
  \citenamefont {Ogata}, \citenamefont {Sakai},\ and\ \citenamefont
  {Awano}}]{Kazuyuki2015}%
  \BibitemOpen
  \bibfield  {author} {\bibinfo {author} {\bibfnamefont {K.}~\bibnamefont
  {Ito}}, \bibinfo {author} {\bibfnamefont {T.}~\bibnamefont {Ogata}}, \bibinfo
  {author} {\bibfnamefont {T.}~\bibnamefont {Sakai}}, \ and\ \bibinfo {author}
  {\bibfnamefont {Y.}~\bibnamefont {Awano}},\ }\href@noop {} {\bibfield
  {journal} {\bibinfo  {journal} {Applied Physics Express}\ }\textbf {\bibinfo
  {volume} {8}},\ \bibinfo {pages} {025101} (\bibinfo {year}
  {2015})}\BibitemShut {NoStop}%
\bibitem [{\citenamefont {Chu}\ and\ \citenamefont {Chen}(2014)}]{Chu2014}%
  \BibitemOpen
  \bibfield  {author} {\bibinfo {author} {\bibfnamefont {T.}~\bibnamefont
  {Chu}}\ and\ \bibinfo {author} {\bibfnamefont {Z.}~\bibnamefont {Chen}},\
  }\href {\doibase 10.1021/nn500043y} {\bibfield  {journal} {\bibinfo
  {journal} {ACS Nano}\ }\textbf {\bibinfo {volume} {8}},\ \bibinfo {pages}
  {3584} (\bibinfo {year} {2014})}\BibitemShut {NoStop}%
\bibitem [{\citenamefont {Tinkham}(1996)}]{Tinkham1996}%
  \BibitemOpen
  \bibfield  {author} {\bibinfo {author} {\bibfnamefont {M.}~\bibnamefont
  {Tinkham}},\ }\href@noop {} {\emph {\bibinfo {title} {Introduction to
  Superconductivity}}}\ (\bibinfo  {publisher} {McGraw-Hill},\ \bibinfo {year}
  {1996})\ pp.\ \bibinfo {pages} {213--230}\BibitemShut {NoStop}%
\bibitem [{\citenamefont {Dynes}\ and\ \citenamefont
  {Fulton}(1971)}]{Dynes1971}%
  \BibitemOpen
  \bibfield  {author} {\bibinfo {author} {\bibfnamefont {R.~C.}\ \bibnamefont
  {Dynes}}\ and\ \bibinfo {author} {\bibfnamefont {T.~A.}\ \bibnamefont
  {Fulton}},\ }\href {\doibase 10.1103/PhysRevB.3.3015} {\bibfield  {journal}
  {\bibinfo  {journal} {Phys. Rev. B}\ }\textbf {\bibinfo {volume} {3}},\
  \bibinfo {pages} {3015} (\bibinfo {year} {1971})}\BibitemShut {NoStop}%
\bibitem [{\citenamefont {Barone}\ and\ \citenamefont
  {Paterno}(1985)}]{BaronePaterno1982}%
  \BibitemOpen
  \bibfield  {author} {\bibinfo {author} {\bibfnamefont {A.}~\bibnamefont
  {Barone}}\ and\ \bibinfo {author} {\bibfnamefont {G.}~\bibnamefont
  {Paterno}},\ }\href@noop {} {\emph {\bibinfo {title} {Physics and
  Applications of the Josephson Effect}}}\ (\bibinfo  {publisher} {Wiley},\
  \bibinfo {year} {1985})\ pp.\ \bibinfo {pages} {70--95}\BibitemShut {NoStop}%
\bibitem [{\citenamefont {Hart}\ \emph {et~al.}(2014)\citenamefont {Hart},
  \citenamefont {Ren}, \citenamefont {Wagner}, \citenamefont {Leubner},
  \citenamefont {Mühlbauer}, \citenamefont {Brüne}, \citenamefont {Buhmann},
  \citenamefont {Laurens W.~Molenkamp},\ and\ \citenamefont
  {Yacoby}}]{Hart2014}%
  \BibitemOpen
  \bibfield  {author} {\bibinfo {author} {\bibfnamefont {S.}~\bibnamefont
  {Hart}}, \bibinfo {author} {\bibfnamefont {H.}~\bibnamefont {Ren}}, \bibinfo
  {author} {\bibfnamefont {T.}~\bibnamefont {Wagner}}, \bibinfo {author}
  {\bibfnamefont {P.}~\bibnamefont {Leubner}}, \bibinfo {author} {\bibfnamefont
  {M.}~\bibnamefont {Mühlbauer}}, \bibinfo {author} {\bibfnamefont
  {C.}~\bibnamefont {Brüne}}, \bibinfo {author} {\bibfnamefont
  {H.}~\bibnamefont {Buhmann}}, \bibinfo {author} {\bibfnamefont {L.~W.}\
  \bibnamefont {Laurens W.~Molenkamp}}, \ and\ \bibinfo {author} {\bibfnamefont
  {A.}~\bibnamefont {Yacoby}},\ }\href@noop {} {\bibfield  {journal} {\bibinfo
  {journal} {Nature Physics}\ }\textbf {\bibinfo {volume} {10}},\ \bibinfo
  {pages} {638} (\bibinfo {year} {2014})}\BibitemShut {NoStop}%
\bibitem [{\citenamefont {Allen}\ \emph {et~al.}(2015)\citenamefont {Allen},
  \citenamefont {Shtanko}, \citenamefont {Fulga}, \citenamefont {Akhmerov},
  \citenamefont {Watanabi}, \citenamefont {Taniguchi}, \citenamefont
  {Jarillo-Herrero}, \citenamefont {Levitov},\ and\ \citenamefont
  {Yacoby}}]{Allen2015}%
  \BibitemOpen
  \bibfield  {author} {\bibinfo {author} {\bibfnamefont {M.~T.}\ \bibnamefont
  {Allen}}, \bibinfo {author} {\bibfnamefont {O.}~\bibnamefont {Shtanko}},
  \bibinfo {author} {\bibfnamefont {I.~C.}\ \bibnamefont {Fulga}}, \bibinfo
  {author} {\bibfnamefont {A.}~\bibnamefont {Akhmerov}}, \bibinfo {author}
  {\bibfnamefont {K.}~\bibnamefont {Watanabi}}, \bibinfo {author}
  {\bibfnamefont {T.}~\bibnamefont {Taniguchi}}, \bibinfo {author}
  {\bibfnamefont {P.}~\bibnamefont {Jarillo-Herrero}}, \bibinfo {author}
  {\bibfnamefont {L.~S.}\ \bibnamefont {Levitov}}, \ and\ \bibinfo {author}
  {\bibfnamefont {A.}~\bibnamefont {Yacoby}},\ }\href {\doibase
  10.1038/nphys3534} {\bibfield  {journal} {\bibinfo  {journal} {Nature
  Physics}\ }\textbf {\bibinfo {volume} {12}},\ \bibinfo {pages} {128}
  (\bibinfo {year} {2015})},\ \Eprint {http://arxiv.org/abs/1504.07630}
  {arXiv:1504.07630} \BibitemShut {NoStop}%
\bibitem [{\citenamefont {Milne}(1961)}]{Milne1961}%
  \BibitemOpen
  \bibfield  {author} {\bibinfo {author} {\bibfnamefont {J.~G.~C.}\
  \bibnamefont {Milne}},\ }\href {\doibase 10.1103/PhysRev.122.387} {\bibfield
  {journal} {\bibinfo  {journal} {Phys. Rev.}\ }\textbf {\bibinfo {volume}
  {122}},\ \bibinfo {pages} {387} (\bibinfo {year} {1961})}\BibitemShut
  {NoStop}%
\bibitem [{\citenamefont {Camara}\ \emph {et~al.}(2008)\citenamefont {Camara},
  \citenamefont {Rius}, \citenamefont {Huntzinger}, \citenamefont {Tiberj},
  \citenamefont {Mestres}, \citenamefont {Godignon},\ and\ \citenamefont
  {Camassel}}]{Camara2008}%
  \BibitemOpen
  \bibfield  {author} {\bibinfo {author} {\bibfnamefont {N.}~\bibnamefont
  {Camara}}, \bibinfo {author} {\bibfnamefont {G.}~\bibnamefont {Rius}},
  \bibinfo {author} {\bibfnamefont {J.~R.}\ \bibnamefont {Huntzinger}},
  \bibinfo {author} {\bibfnamefont {A.}~\bibnamefont {Tiberj}}, \bibinfo
  {author} {\bibfnamefont {N.}~\bibnamefont {Mestres}}, \bibinfo {author}
  {\bibfnamefont {P.}~\bibnamefont {Godignon}}, \ and\ \bibinfo {author}
  {\bibfnamefont {J.}~\bibnamefont {Camassel}},\ }\href {\doibase
  10.1063/1.2988645} {\bibfield  {journal} {\bibinfo  {journal} {Applied
  Physics Letters}\ }\textbf {\bibinfo {volume} {93}},\ \bibinfo {pages} {2008}
  (\bibinfo {year} {2008})},\ \Eprint {http://arxiv.org/abs/0806.4056}
  {0806.4056} \BibitemShut {NoStop}%
\bibitem [{\citenamefont {Sprinkle}\ \emph {et~al.}(2010)\citenamefont
  {Sprinkle}, \citenamefont {Ruan}, \citenamefont {Hu}, \citenamefont
  {Hankinson}, \citenamefont {Rubio-Roy}, \citenamefont {Zhang}, \citenamefont
  {Wu}, \citenamefont {Berger},\ and\ \citenamefont {de~Heer}}]{Sprinkle2010}%
  \BibitemOpen
  \bibfield  {author} {\bibinfo {author} {\bibfnamefont {M.}~\bibnamefont
  {Sprinkle}}, \bibinfo {author} {\bibfnamefont {M.}~\bibnamefont {Ruan}},
  \bibinfo {author} {\bibfnamefont {Y.}~\bibnamefont {Hu}}, \bibinfo {author}
  {\bibfnamefont {J.}~\bibnamefont {Hankinson}}, \bibinfo {author}
  {\bibfnamefont {M.}~\bibnamefont {Rubio-Roy}}, \bibinfo {author}
  {\bibfnamefont {B.}~\bibnamefont {Zhang}}, \bibinfo {author} {\bibfnamefont
  {X.}~\bibnamefont {Wu}}, \bibinfo {author} {\bibfnamefont {C.}~\bibnamefont
  {Berger}}, \ and\ \bibinfo {author} {\bibfnamefont {W.~a.}\ \bibnamefont
  {de~Heer}},\ }\href {\doibase 10.1038/nnano.2010.192} {\bibfield  {journal}
  {\bibinfo  {journal} {Nature nanotechnology}\ }\textbf {\bibinfo {volume}
  {5}},\ \bibinfo {pages} {727} (\bibinfo {year} {2010})}\BibitemShut {NoStop}%
\end{thebibliography}%

\end{document}